\newcommand{\diracslash}[1]{#1\llap{/\kern2pt}}

\newcommand{\be}{\begin{equation}}
\newcommand{\ee}{\end{equation}}
\newcommand{\bea}{\begin{eqnarray}}
\newcommand{\eea}{\end{eqnarray}}
\newcommand{\ba}[1]{\begin{array}{#1}}
\newcommand{\ea}{\end{array}}

\newcommand{\bt}{\begin{tabular}}
\newcommand{\et}{\end{tabular}}

\newcommand{\beas}{\begin{eqnarray*}}
\newcommand{\eeas}{\end{eqnarray*}}

\RequirePackage{lineno}

\documentclass[preprint,prd,aps,floats,nofootinbib,floatfix]
 {revtex4-2}

\DeclareSymbolFont{rsfs}{U}{rsfs}{m}{n}
\DeclareSymbolFontAlphabet{\mathrsfs}{rsfs}

\usepackage{amsmath}
\usepackage{graphicx}
\usepackage[mathscr]{eucal}
\usepackage{multirow}
\usepackage{graphicx,epstopdf}
\usepackage{graphicx}
\usepackage{subcaption}
\usepackage{cancel}

\usepackage{setspace}
\usepackage[utf8]{inputenc}
\usepackage{csquotes}
\usepackage[english]{babel}
\usepackage{hyperref}
\usepackage{upgreek}
\hypersetup{
    colorlinks=true,
    linkcolor=blue,
    filecolor=magenta,     
    citecolor=blue, 
    urlcolor=blue,
}
\usepackage{booktabs} 
\usepackage{cleveref}
\usepackage{caption}
\usepackage{rotating}

\begin{document}
\setstretch{1.5}
%\begin{linenumbers}
%\begin{linenumbers}
\title{Effect of dark matter and scalar-cut potential on the macroscopic properties of neutron stars: A chiral quark mean field approach} 
%%%%%%%%%%%%%%%%%%%%%%%%%%%%%%%%%%%%%%%%%%%%%%%%%%%%%%%%%%%%%%%%%
\author{Hrishika P}
\email{hrishikap2003@gmail.com}
\author{Arvind Kumar}
\email{kumara@nitj.ac.in}
\affiliation{Department of Physics, Dr. B R Ambedkar National Institute of Technology Jalandhar, 
	Jalandhar -- 144008, Punjab, India}

\def\be{\begin{equation}}
\def\ee{\end{equation}}
\def\bearr{\begin{eqnarray}}
\def\eearr{\end{eqnarray}}
\def\zbf#1{{\bf {#1}}}
\def\bfm#1{\mbox{\boldmath $#1$}}
\def\hf{\frac{1}{2}}
\def\kp{\zbf k+\frac{\zbf q}{2}}
\def\km{-\zbf k+\frac{\zbf q}{2}}
\def\hwo{\hat\omega_1}
\def\hwt{\hat\omega_2}

\begin{abstract}

We investigate the effects of fermionic dark matter (DM) and scalar-cut potential ($U_{\text{cut}}(\phi)$) on
the equation of state (EOS) and macroscopic properties of neutron stars. The baryonic matter consists of neutrons, protons, electrons, and muons under the $\beta$-equilibrium and is described within the chiral SU(3) quark mean-field (CQMF) model. In this framework, quarks are confined within baryons, whose properties are modified through the exchange of scalar fields $\sigma$, $\zeta$, and $\delta$, and vector fields $\omega$ and $\rho$. The DM is
modelled as a self-interacting fermionic fluid with scalar $\varphi_D$ and vector $V_D^\mu$ mediators. The baryonic and DM fluids interact only through gravity and are treated within a two-fluid Tolman–Oppenheimer–Volkoff formalism. We examine the resulting mass–radius relation and tidal deformability, along with the gravitational redshift, to investigate the effects of DM on neutron star structure and observable properties. The inclusion of $U_{\text{cut}}(\phi)$ stiffens the EOS, whereas increasing the DM fraction generally softens the EOS and modifies the mass–radius relation, tidal deformability, and gravitational redshift. The magnitude and nature of these modifications depend on the DM particle mass and mediator interactions, which determine whether the resulting configurations exhibit compact DM cores or extended DM halos.
\end{abstract}

\maketitle

\newpage

\section{Introduction}
\label{intro}Dark matter (DM) is one of the most compelling unresolved components of the
Universe. Its existence is supported by a broad range of gravitational
observations, including the anomalous rotation of galaxies, gravitational
lensing, the dynamics of galaxy clusters, the formation of large scale
structures, and the anisotropies of the cosmic microwave background~\cite{Rubin:1980,Bosma:1981,Clowe:2006,Springel:2005,Ade:2016}. In the
standard $\Lambda$ Cold Dark Matter ($\Lambda$CDM) cosmological model, DM contributes approximately
27\% of the total energy density of the Universe, substantially exceeding
the contribution from ordinary baryonic matter~\cite{Aghanim:2020}. Despite extensive efforts in direct detection experiments, no conclusive
particle level detection of DM has been established
\cite{Aprile:2018,Aprile:2025,Aalbers:2025,Bo:2025}. This persistent absence of a confirmed signal has motivated the exploration of alternative approaches that rely on the gravitational effects of DM rather than its direct interaction with ordinary matter. In particular, compact astrophysical objects such as neutron stars and white dwarfs provide natural environments in which DM can be captured and accumulated over long timescales~\cite{Kouvaris:2008,Issifu:2025,Lopes:2025,deLavallaz:2010}. 
% Their strong gravitational fields and extreme densities make neutron stars particularly promising laboratories for studying DM-induced changes in stellar structure and observables.

Neutron stars are the compact remnants of massive stars and represent some of the densest stable objects in the Universe. Their interiors can reach several times the nuclear saturation density, providing a unique environment for studying cold, ultra dense matter under conditions that are inaccessible to terrestrial experiments~\cite{Lattimer:2004,Ozel:2016,Baym:2018,Lattimer:2021}. 
The extreme gravitational fields of neutron stars also makes them promising laboratories for investigating DM. DM particles passing through a neutron star can lose kinetic energy through scattering with baryonic matter (BM) and become gravitationally bound to the star~\cite{Busoni:2021,Bell:2020}. Repeated scattering can then lead to the thermalization and concentration of captured DM toward the stellar interior~\cite{garani:19,Bertoni:2013,Bell:2023}. The amount of accumulated DM depends on several factors, including the DM mass, interaction strength, ambient DM density, and the age and astrophysical environment of the neutron star~\cite{Bramante:2024}. The presence of DM within neutron stars can influence their structural properties such as mass-radius ($M-R$) relationship. The tidal deformability $\Lambda$ is the measure of deformation of stellar structure due to gravitational field of its companion star and $\Lambda$ can be affected by the presence of DM. In general, neutron stars have masses of about 1.4–2.5$M_\odot$, radii of $10–15$ km, and $\Lambda$ of roughly $100–1000$~\cite{Chandrasekhar:1939,Nattila:2022}. Depending on the properties and abundance of DM, it may form either a compact core or an extended halo within the neutron star, leading to a softer or stiffer equation of state (EOS), respectively~\cite{Das:2019} creating a dark matter admixed neutron star (DANS).

% Dark matter (DM) can accumulate inside neutron stars (neutron stars) through capture~\cite{Goldman:1989,Kouvaris:2008,Kouvaris:2010,deLavallaz:2010,Raj:2018}, and, for asymmetric self-interacting DM~\cite{Zurek:2014,Tulin:2018}, through gravothermal collapse into compact dark objects that subsequently accrete visible matter~\cite{Kouvaris:2015,Ellis:2018}; comprehensive reviews are given in Refs.~\cite{Bramante:2024,Grippa:2025}. 

The resulting DANS depend strongly on the assumed DM candidate. There are fermionic DM, coupled to nucleons via the Higgs portal, which behaves effectively as a single fluid has been studied in relativistic mean-field (RMF) models~\cite{Panotopoulos:2017,Das:2019,Das:2022Galaxies,Lourenco:2022,Kumar:2024}, Bosonic DM~\cite{Ellis:2018,Sagun:2023,Karkevandi:2022,Karkevandi:2024}, mirror DM~\cite{Sandin:2009,Ciarcelluti:2011,Hippert:2023}, DM produced through neutron decay~\cite{Fornal:2018,Baym:2018,McKeen:2018,Motta:2018} and asymmetric self-interacting fermionic DM mediated by light bosons has also been constrained using neutron star observations~\cite{Nelson:2019,Ivanytskyi:2020,Shawqi:2024,Rutherford:2024,Biesdorf:2024}. When DM interacts with BM only gravitationally, the star is described by a two-fluid system of coupled Tolman-Oppenheimer-Volkoff (TOV) equations~\cite{Sandin:2009,Ciarcelluti:2011,Leung:2011}, and the tidal deformability~\cite{Hinderer:2008,Hinderer:2010} must be generalised accordingly~\cite{Das:2022,Leung:2022}. Within this framework, a self-interacting fermionic DM with scalar and vector dark mediators in the mean-field approximation was introduced in Ref.~\cite{Xiang:2014}. It was combined with an RMF (FSU2R) baryonic EOS and Bayesian parameter inference in Ref.~\cite{Das:2022}. It was further extended to BSk22, MPA1 and APR4 baryonic EOSs with both linear and quadratic scalar couplings in Ref.~\cite{Grippa:2026}. A vector mediated asymmetric DM model, with DM core and halo configurations was also studied~\cite{Kumar:2025}. 

Existing studies of nuclear matter are primarily based on RMF models or parametrized EOSs~\cite{Das:2022,Grippa:2026}. In these approaches the baryons are treated as point-like particles, so the effect of their quark substructure on the baryonic EOS is not included. To the best of our knowledge, a chiral quark-level approach for describing baryonic matter has not yet been incorporated into a two-fluid DM framework. In the present work we model the BM using the chiral SU(3) quark mean-field (CQMF) model~\cite{Wang:2001,Wang:2005}, where baryons acquire their masses and interactions through confined quarks coupled to meson fields through chirally invariant interactions. To stiffen the EOS at high density, we include the scalar-cut potential ($U_{\rm cut}(\phi)$)~\cite{Maslov:2015a,Maslov:2015,Maslov:2016,Kolomeitsev:2017}. We couple it gravitationally, to fermionic DM interacting via scalar and vector mediators, and we compare neutron star properties with and without $U_{\rm cut}(\phi)$, both in the absence and presence of DM, focusing on the $M-R$ and $\Lambda-M$ relations and the gravitational redshift, to investigate the combined effects of the high density modification of the baryonic EOS and DM interactions on neutron star structure. Our study thus provides a framework that can be made consistent with massive pulsar observations and explores how DM properties influence these observables.

The structure of the paper is as follows. Section~\ref{sec:methodology_BM} introduces the CQMF model, detailing the associated field equations and density formulas used to describe BM. Section~\ref{sec:methodology_DM} describes the fermionic DM with scalar and vector interactions, while Sec.~\ref{sec:methodology_DANS} presents the two-fluid formalism used to construct DANS configurations. Section~\ref{sec:results} discusses the resulting neutron star properties, including the $M-R$ and $\Lambda-M$ relation, radial pressure profiles, and gravitational redshift, with a systematic comparison of the results obtained with and without the $U_{\text{cut}}(\phi)$. Finally, Sec.~\ref{sec:conclusion} summarizes the main findings and conclusions of this work.

\section{Methodology}
\label{sec:methodology}

\subsection{Baryonic matter (BM)}
\label{sec:methodology_BM}
In the present study, the BM for the neutron star is formulated using the chiral SU(3) quark mean-field model (CQMF) framework, where the constituent quarks are confined inside baryons through an effective confining potential~\cite{Wang:2001npa,Papazoglou:1999prc}.
The effective Lagrangian density  is given as~\cite{Papazoglou:1999prc,Mishra2024Charmonium}
\setcounter{equation}{1}
\begin{align}
 \mathcal{L}_{\text{eff}}= \mathcal{L}_{\text{q0}} + \mathcal{L}_{qm} + \mathcal{L}_{\Sigma \Sigma} + \mathcal{L}_{\text{VV}} + \mathcal{L}_{{\chi}SB} + \mathcal{L}_{\Delta m}+\mathcal{L}_{\text{c}}. \label{lagrangian}
\end{align}
Here, $\mathcal{L}_{\text{q0}}= \bar{q}i\gamma^{\mu}\partial_{\mu}q$ represents the kinetic contribution of the massless quarks, while $\mathcal{L}_{qm}$ denotes the invariant quark-meson interaction term~\cite{Ping:2001ctp,Singh:2017cpc} and is given by
\begin{equation}
\mathcal{L}_{qm}=
g_s\left(\bar{q}_L M q_R+\bar{q}_R M^\dagger q_L\right)-
g_v\left(\bar{q}_L\gamma^\mu l_\mu q_L+\bar{q}_R\gamma^\mu r_\mu q_R\right),
\label{eq:L_qm}
\end{equation}
where $q_L$ and $q_R$ denote the left and right handed quark fields, and $g_s$ and $g_v$ denote the corresponding scalar and vector coupling constant, respectively.
The third and fourth terms, $\mathcal{L}_{\Sigma\Sigma}$ and $\mathcal{L}_{VV}$
are the chiral invariant scalar and vector meson self-interaction terms, respectively, given by 
\begin{align}
\mathcal{L}_{\Sigma\Sigma} ={}&
-\frac{1}{2}k_0\chi^2(\sigma^2+\zeta^2+\delta^2)
+k_1(\sigma^2+\zeta^2+\delta^2)^2
+k_2\left(\frac{\sigma^4}{2}+\frac{\delta^4}{2}
+3\sigma^2\delta^2+\zeta^4\right)
\nonumber\\
&+k_3\chi(\sigma^2-\delta^2)\zeta-k_4\chi^4
-\frac{1}{4}\chi^4\ln\frac{\chi^4}{\chi_0^4}
+\frac{d}{3}\chi^4
\ln\left(\frac{(\sigma^2-\delta^2)\zeta}
{\sigma_0^2\zeta_0}
\right),
\label{eqn2}
\end{align}
and
\begin{equation}
\mathcal{L}_{VV}
=
\frac{1}{2}\frac{\chi^2}{\chi_0^2}
\left(m_\omega^2\omega^2+m_\rho^2\rho^2\right)
+g_4\left(\omega^4+6\omega^2\rho^2+\rho^4\right).
\end{equation}
In Eq.(\ref{eqn2}), the last two logarithmic terms are introduced to account for the scale breaking effect and to evaluate the trace of the energy-momentum tensor. Also, $\sigma_0$, $\zeta_0$ and $\chi_0$ represents the vacuum expectation values of scalar fields $\sigma$, $\zeta$ and the dilaton field $\chi$, respectively. They are expressed as
\begin{equation}
    \sigma_0=-F_\pi, \qquad    \zeta_0=\frac{1}{\sqrt2}(F_\pi-2F_K)
\end{equation}
and $\chi_0$ is given in Table~\ref{params}. The explicit chiral symmetry breaking is incorporated through the last three terms of Eq. (\ref{lagrangian}). The term, $\mathcal{L}_{{\chi}SB}$ generates non-vanishing masses for pseudoscalar
mesons and the explicit strange quark mass contribution is introduced through $\mathcal{L}_{\Delta m}$. 
Whereas the last term $\mathcal{L}_{c}$ represents the confinement of constituent quarks inside the baryon.

The Dirac equation describing a quark field $\Psi_{ij}$ in the
presence of the meson mean-fields can be written as
\begin{equation}
\left[-i\vec{\alpha}\cdot\vec{\nabla}
+\beta\chi_c(r)+\beta m_i^*\right]\Psi_{ij}
=e_i^*\Psi_{ij},
\end{equation}
where, $\vec{\alpha}=\gamma^0\vec{\gamma}$ and $\beta=\gamma^0$.
Here, the indices $i$ and $j$ represent the quarks ($i=u,d$ and $s$) and the baryons ($j=n$ and $p$),
respectively. The quantity $\chi_c(r)$ denotes the confining
potential, which provides quark confinement through the meson fields. It is given by
$\chi_c=\frac{1}{4}k_c r^2 (1+\gamma^0)$, where coupling constant $k_c$ is taken to be $300$ MeV $\rm fm^{-2}$. The effective quark mass $m_i^*$ and the corresponding
effective quark energy $e_i^*$ are defined as
\begin{equation}
m_i^*=-g_{\sigma i}\sigma-g_{\zeta i}\zeta
-g_{\delta i}\tau_3\delta,
\end{equation}
and
% \begin{equation}
% e_i^*=m_i^*
% +\frac{3\sqrt{k_c}}
% {\sqrt{2(e_i^*+m_i^*)}}.
% \end{equation}
\begin{equation}
e_i^*=e_i-g_{\omega i}\omega-g_{\rho i}\tau_{3i}\rho.
\end{equation}
Here, $g_{\sigma i}$, $g_{\zeta i}$, and $g_{\delta i}$ denote the coupling constants of the scalar fields $\sigma$, $\zeta$, and $\delta$, respectively, while $g_{\omega i}$ and $g_{\rho i}$ denote the coupling constants of the vector fields $\omega$ and $\rho$, respectively.
To obtain massive neutron star configurations, we introduce scalar-cut potential ($U_{\text{cut}}(\phi)$) that suppresses the monotonic decrease of the scalar field with increasing nucleon density, resulting in a stiffer EOS, given by~\cite{Maslov:2015,Pthakur:2024,Zhang:2018}
\begin{equation}
U_{\mathrm{cut}}(\phi)
=
\alpha
\ln\left\{
1+\exp\left[
\bar\beta
\left(
1-\frac{m_i^*}{m_i}-f_s
\right)
\right]
\right\},
\label{Eq:cutpotential}
\end{equation}
where $\phi=(\sigma, \zeta, \text{and }\delta)$, $\alpha=m_\pi^4$ and $\bar\beta=120$ are fixed parameters, while $f_s$ is a free parameter controlling the strength of the $U_{\text{cut}}(\phi)$~\cite{Maslov:2015}. At zero temperature, the thermodynamic potential can be written as
\begin{equation}
\begin{aligned}
\Omega ={}&\Omega_j+\Omega_l=
-\sum_{j=n,p} \frac{\gamma_j}{48\pi^2}
\Bigg\{
E_{F_j}^{*}
\left(2k_{F_j}^{3}-3M_j^{*2}k_{F_j}\right)
+3M_j^{*4}
\ln\left(
\frac{k_{F_j}+E_{F_j}^{*}}{M_j^*}
\right)
\Bigg\}
\\
&-\sum_{l=e,\mu} \frac{\gamma_l}{48\pi^2}
\Bigg\{
E_{F_l}
\left(2k_{F_l}^{3}-3m_l^2 k_{F_l}\right)
+3m_l^4
\ln\left(
\frac{k_{F_l}+E_{F_l}}{m_l}
\right)
\Bigg\}
-\mathcal{L}'.
\end{aligned}
\label{thermo_pot}
\end{equation}
The last term $\mathcal{L}'$ is defined as
$\mathcal{L'} = \mathcal{L}_{M}-\mathcal{L}_{vac}-U_{\text{cut}}(\phi)$ where $\mathcal{L}_{M}=\mathcal{L}_{VV} + \mathcal{L}_{\Sigma\Sigma}
+ \mathcal{L}_{\chi SB}$. Here, $\mathcal{L}_{vac}$ is introduced to account for the vacuum energy associated with the meson fields and maintain consistency with the vacuum mass calibration of baryons. The quantities $k_{F_j}$  and $k_{F_l}$ represents the Fermi momenta of the baryons and leptons, respectively, and are given by
\begin{equation}
k_{F_j}=\sqrt{\nu_j^{*2}-M_j^{*2}}
\qquad \text{and}\ \qquad
k_{F_l}=\sqrt{\nu_l^2-m_l^2}.
\end{equation}
Here, $\nu_j^*$ is the effective chemical potential of the baryon, given by $\nu_j^*=\nu_j-g_{\omega j}\omega
-g_{\rho j}\tau_{3j}\rho$,
and  $\nu_l$ and $m_l$ denote the chemical potential and bare mass of the lepton, respectively. Since leptons do not couple to the meson fields, their masses and chemical potentials remain unmodified. The effective baryon mass $M_j^*$, is determined by the scalar mean fields of the model and is given by
\begin{equation}
M_j^* =
\sqrt{
\left(
\sum_i n_{ij} e_i^* + E_j^{\mathrm{spin}}
\right)^2-
\left\langle p_{\mathrm{j,cm}}^{*2}\right\rangle
}.
\end{equation}
Here, $n_{ij}$ is the number of quarks $i$ in a baryon $j$ and \(\langle p_{j,\mathrm{cm}}^{2} \rangle\) represents the correction arising from the
spurious center of mass motion of the constituent quarks~\cite{Barik:2013prc,Barik:1985prd}. To determine the scalar fields $\sigma$, $\zeta$ and $\delta$, the dilaton field $\chi$ and vector fields $\omega$ and $\rho$, the total thermodynamic potential is minimized with respect to the meson fields~\cite{Kumari:2020mci,Kumar:2019cpc,Kumar:2019epjc} such that,
\begin{equation}
\frac{\partial \Omega}{\partial \sigma}
=
\frac{\partial \Omega}{\partial \zeta}
=
\frac{\partial \Omega}{\partial \delta}
=
\frac{\partial \Omega}{\partial \chi}
=
\frac{\partial \Omega}{\partial \omega}
=
\frac{\partial \Omega}{\partial \rho}
=0 .
\label{Eq:mean_fields}
\end{equation}
The above conditions yield a set of coupled nonlinear equations for the mean fields (see Appendix), which are solved self-consistently together with the conditions governing the composition of neutron star matter, determined by imposing baryon number conservation, charge neutrality, and $\beta$-equilibrium. At a given baryon density $\rho_B$, these conditions are expressed as
\begin{equation}
\rho_B=\rho_v^p+\rho_v^n,
\qquad
\rho_v^p=\rho_v^e+\rho_v^\mu,
\qquad
\nu_n-\nu_p=\nu_e=\nu_\mu,
\label{eq:equilibrium_conditions}
\end{equation}
where $\rho_v^i$ denote the vector density of the particle species. The resulting field equations depend on the scalar $\rho_s^i$ and vector densities $\rho_v^i$ of the baryons. For a baryon $j$, these densities are defined as

\begin{align}
\rho_s^j
&=
\frac{M_j^*}{\pi^2}
\int_0^{k_{F_j}}
\frac{k^2\,dk}
{\sqrt{M_j^{*2}+k^2}}
\nonumber\\[4pt]
&=
\frac{M_j^{*3}}{2\pi^2}
\left[
\frac{k_{F_j}}{M_j^*}
\sqrt{1+\frac{k_{F_j}^2}{M_j^{*2}}}
-
\ln\left(
\frac{k_{F_j}}{M_j^*}
+
\sqrt{1+\frac{k_{F_j}^2}{M_j^{*2}}}
\right)
\right]
\end{align}
and
\begin{align}
\rho_v^j
&=
\frac{\gamma_j}{2\pi^2}
\int_0^{k_{F_j}} k^2\,dk = 
\frac{\gamma_j k_{F_j}^3}{6\pi^2}.
\label{Eq_bm: rho_v}
\end{align}
Using the thermodynamic potential obtained from Eq. (\ref{thermo_pot}) and the baryon number density given by Eq. (\ref{Eq_bm: rho_v}), the energy density $\epsilon_{BM}$ and pressure $P_{BM}$ of the baryonic system can be obtained as
\begin{equation}
P_{BM}=-\Omega \qquad
\epsilon_{BM} =
\Omega
+\sum_{j} \nu_j\rho_v^j+\sum_{l} \nu_l\rho_v^l.
\label{Eq:bm_pressure}
\end{equation}
These expressions provide the thermodynamic quantities required to construct the EOS of the BM. The resulting EOS is then used to investigate the properties of neutron stars under different physical conditions.

\subsection{Dark matter (DM)}
\label{sec:methodology_DM}
We model the dark component as a self interacting fermionic DM fluid which are assumed to interact through the exchange of scalar ($\varphi_D$) and vector ($V^{\mu}_D$) mediator fields~\cite{Xiang:2014,Pthakur:2024,Grippa:2026}. The scalar interaction modifies the effective mass of the dark fermion, whereas the vector interaction contributes a repulsive component to the energy density and pressure of the DM. This framework provides a thermodynamically consistent EOS for the DM component. The Lagrangian for the DM fluid is given by~\cite{Xiang:2014}
\begin{equation}
\begin{aligned}
\mathcal{L}_{\rm DM} ={}&
{\bar\chi}
\left[
\gamma^\mu
\left(i\partial_\mu-g_{V_D} V_{D\mu}\right)
-\left(M_\chi-g_{\varphi_D}\varphi_D\right)
\right]\chi
+\frac{1}{2}
\left(
\partial_\mu\varphi_D\,\partial^\mu\varphi_D
-m_{\varphi^2_{D}}\varphi_D^2
\right)\\
&
-\frac{1}{4}V_{D\mu\nu}V_D^{\mu\nu}
+\frac{1}{2}m_{V_D}^2V_{D\mu} V_D^\mu,
\end{aligned}
\label{dm_lagrangian}
\end{equation}
where, $M_\chi$ is the mass of the DM fermion, whereas $m_{\varphi_{D}}$ and $m_{V_D}$ denote the
masses of the dark scalar and vector mesons, respectively. The term $V_{\mu\nu}$ represents the field strength tensor of the vector DM field, which is given by $V_{\mu\nu}=\partial_\mu V_{D\nu}-\partial_\nu V_{D\mu}$. And $g_{V_D}$ and $g_{\varphi_D}$ denote the coupling constants associated with the neutral vector and scalar fields, respectively, which describe their interactions with the DM fermion $\chi$ and are defined as~\cite{Grippa:2026} 
\begin{equation}
    g_{V_D}=\frac{d_{V_D}M_\chi}{M_p},\qquad \text{and} \qquad g_{\varphi_D}=\frac{d_{\varphi_D} M_\chi}{M_p}.
\end{equation}
Here, $M_p$ is the reduced Planck mass and we assume $d_{V_D}$ and $d_{\varphi_D}$ as free, dimensionless parameters. 

Using the Lagrangian in Eq. (\ref{dm_lagrangian}), the equations of motion governing the DM fermion and the scalar and vector mediator fields can be derived through the Euler-Lagrange equations. These equations are given by~\cite{Grippa:2026}
\begin{equation}
\left[
\gamma^\mu\left(i\partial_\mu-g_{V_D}V_{D\mu}\right)
-\left(M_\chi-g_{\varphi_D}\varphi_D\right)
\right]\chi=0,
\end{equation}
\begin{equation}
\partial_\mu\partial^\mu\varphi_D
+m_{\varphi_D}^2\varphi_D
=
g_{\varphi_D}\,\bar{\chi}\chi,
\end{equation}
\begin{equation}
\partial_\mu V_D^{\mu\nu}
+m_{V_D}^2V_D^\mu
=
g_{V_D}\,\bar{\chi}\gamma^\mu\chi.
\end{equation}
Within the mean-field approximation, the scalar and vector mediator fields are replaced by their corresponding ground state expectation values. For uniform and isotropic DM matter, the spatial components of the vector field vanish, leaving only the temporal component. Therefore the resulting field operators $\varphi_{D}^0$ and $V_D^0$ are given by
\begin{equation}
\varphi_{D}^0
=
\frac{g_{\varphi_D}}{m_{\varphi_D}^2}
\langle\bar{\chi}\chi\rangle
=
\frac{g_{\varphi_D}}{m_{\varphi_D}^2}
\rho_s^\chi, \quad \text{and}\quad  V_D^0
=
\frac{g_{V_D}}{m_{V_D}^2}
\langle\chi^\dagger\chi\rangle=\frac{g_{V_D}}{m_{V_D}^2}\rho_v^\chi,
\label{scalar0}
\end{equation}
where $\rho_s^\chi$ and $\rho_v^\chi$ are the scalar and vector densities of DM fluid which are expressed as 
\begin{equation}
\rho_s^\chi
=
\frac{M_\chi^{*3}}{2\pi^2}
\left[
\frac{k_{F_\chi}}{M_\chi^*}
\sqrt{1+\left(\frac{k_{F_\chi}}{M_\chi^*}\right)^2}
-
\ln\left(
\frac{k_{F_\chi}}{M_\chi^*}
+
\sqrt{1+\left(\frac{k_{F_\chi}}{M_\chi^*}\right)^2}
\right)
\right],
\label{scalar_density_dm}
\end{equation}
and
\begin{equation}
\rho_v^\chi=
\frac{\gamma_\chi k_{F_\chi}^3}{6\pi^2}.
\label{vector0}
\end{equation}
respectively. Where, $\gamma_\chi=2$, $k_{F_\chi}$ is the DM Fermi momentum and $M_\chi^*$ is the effective mass of DM, given by
\begin{equation}
M_\chi^*=M_\chi-g_{\varphi_D}\varphi_D^0 .
\end{equation}
The energy density $\epsilon_{DM}$ and pressure $P_{DM}$ is obtained by calculating the energy momentum tensor with 
Lagrangian density given in Eq.(\ref{dm_lagrangian}), and are given by~\cite{Xiang:2014}
\begin{equation}
\epsilon_{\rm DM}
=
\epsilon_\chi^{\rm kin}
+
\frac{1}{2}m_{V_D}^2V_D^{02}
+
\frac{1}{2}m_{\varphi_D}^2\varphi_D^{02} ,
\end{equation}
\begin{equation}
P_{\rm DM}
=
P_\chi^{\rm kin}
+
\frac{1}{2}m_{V_D}^2V_D^{02}
-
\frac{1}{2}m_{\varphi_D}^2\varphi_D^{02} .
\end{equation}
Here, $\epsilon_\chi^{\rm kin}$ and $P_\chi^{\rm kin}$ represent the
kinetic contributions to the energy density and pressure, respectively.
The kinetic energy density is given by~\cite{narain:2006}
\begin{equation}
\begin{aligned}
\epsilon_\chi^{\rm kin}
&=
\frac{\gamma_\chi}{2\pi^2}
\int_0^{k_{F_\chi}}
dk\,k^2
\sqrt{k^2+M_\chi^{*2}}
\\
&=
\frac{1}{8\pi^2}
\left[
k_{F_\chi}E_{F_\chi}^*
\left(2k_{F_\chi}^2+M_\chi^{*2}\right)
-
M_\chi^{*4}
\ln\left(
\frac{k_{F_\chi}+E_{F_\chi}^*}
{M_\chi^*}
\right)
\right],
\end{aligned}
\label{E_kin}
\end{equation}
where the effective Fermi energy is defined as $E_{F_\chi}^*=\sqrt{k_{F_\chi}^2+M_\chi^{*2}}$.
Similarly, the kinetic pressure is given by~\cite{narain:2006}
\begin{equation}
\begin{aligned}
P_\chi^{\rm kin}
&=
\frac{\gamma_\chi}{6\pi^2}
\int_0^{k_{F_\chi}}
dk\,
\frac{k^4}
{\sqrt{k^2+M_\chi^{*2}}}
\\
&=
\frac{1}{24\pi^2}
\left[
k_{F_\chi}E_{F_\chi}^*
\left(2k_{F_\chi}^2-3M_\chi^{*2}\right)
+
3M_\chi^{*4}
\ln\left(
\frac{k_{F_\chi}+E_{F_\chi}^*}
{M_\chi^*}
\right)
\right].
\end{aligned}
\label{P_kin}
\end{equation}
These resulting EOS is employed to solve the TOV equations and hence to investigate the properties of neutron stars in the following section.

\section{STRUCTURAL PROPERTIES OF DARK MATTER
ADMIXED NEUTRON STAR (DANS)}
\label{sec:methodology_DANS}

The EOS obtained in the previous Secs.~\ref{sec:methodology_BM} and \ref{sec:methodology_DM} provide the
thermodynamic relation between the pressure and energy density of the BM and 
DM, respectively. In order to get the description of DANS with mass $M$, it is needed to solve the widely known TOV equation. In this work, we treat BM and DM as two distinct fluids. Therefore, the TOV equations governing the behaviour of this two-fluid system are given by ~\cite{Das:2022,Rutherford:2023} (units are chosen such that G$ = $c$ =1$)

\label{sec:tov}
\begin{equation}
\frac{dP_i}{dr}
=
-\left(P_i+\epsilon_i\right)
\frac{4\pi r^3 P_{\mathrm{tot}}+M(r)}
{r\left(r-2M(r)\right)},
\label{Eq:tov:P}
\end{equation}

\begin{equation}
\frac{dM(r)}{dr}
=
4\pi \epsilon_{\mathrm{tot}} r^2,
\label{Eq:tov:M}
\end{equation}
where $P_{tot}= P_{BM}+ P_{DM}$, $\epsilon_{tot}= \epsilon_{BM}+\epsilon_{DM}$ are the total pressure and energy density respectively. For each component, the EOS
is specified by
$P_i = P_i(\epsilon_i)$, where $P_i$ and $\epsilon_i$ denote the pressure and energy density of the
corresponding fluid ($i$=BM,DM), respectively. The coupled TOV equations, together with the EOS for
the BM and DM components, are solved numerically by
specifying the central energy density of the star, $\epsilon_c$, within the
relevant range of central densities. The equations are integrated
outward from the stellar center to the surface of the star. For a
two-fluid system, the stellar radius is determined by the
condition that the total pressure vanishes, $P_{\rm tot}(R)=0$.
Along with central density, we need to fix an additional parameter to characterize the relative contribution of dark matter. We therefore specify the DM fraction
$f_{\rm DM}$, defined as
\begin{equation}
f_{\rm DM}
=
\frac{M_{\rm DM}}{M_{\rm tot}},
\end{equation}
where $M_{\rm DM}$ and $M_{\rm tot}$ denote the DM mass and
the total gravitational mass of the star, respectively.

In addition to the mass and radius, the tidal deformability of a
neutron star is an important quantity characterizing its internal
structure. During the late inspiral phase of a binary neutron star
system, the tidal gravitational field generated by each companion
induces a quadrupolar deformation in the other star. The response of
the neutron star to this external tidal field is quantified by its
tidal deformability, which measures the degree to which the star is
deformed under the influence of the companion's gravitational field. The quadrupolar tidal deformability $\lambda$ is related to the
external tidal field $\mathcal{E}_{ij}$ and the induced quadrupole
moment $Q_{ij}$ through
\begin{equation}
Q_{ij}=-\lambda\,\mathcal{E}_{ij},
\end{equation}
where $\lambda$ characterizes the linear response of the neutron star
to the applied tidal field. The dimensionless tidal deformability $\Lambda$ is defined in terms of
the dimensional tidal deformability $\lambda$ as
\begin{equation}
\Lambda
=
\frac{\lambda}{M^5}
=
\frac{2}{3}k_2
\left(\frac{R}{M}\right)^5.
\label{Eq:Lambda}
\end{equation}
Here, $M$ and $R$ denote the gravitational mass and radius of the
neutron star, respectively, and $k_2$ is the quadrupolar tidal Love
number given by~\cite{Flanagan:2008,Damour:2009,Hinderer:2009}
\begin{equation}
\begin{aligned}
k_2={}&
\frac{8C^5}{5}(1-2C)^2
\Bigg[
2+2C(y_R-1)-y_R
\Big\{
2C(6-3y_R+3C(5y_R-8))
\\
&\quad
+4C^3\left(13-11y_R\right)
+C^2(3y_R-2)
+2C^2(1+y_R)
\\
&\quad
+3(1-2C)^2
\left[
2-y_R+2C(y_R-1)
\right]
\ln(1-2C)
\Big\}
\Bigg]^{-1}.
\end{aligned}
\end{equation}
In the above equation, $C= M/R$ is the compactness parameter and
$y_R= y(r=R)$ is the solution of the two-fluid differential equation, which is defined as \cite{Das:2022}
\begin{equation}
r\frac{dy(r)}{dr}
+y(r)^2+y(r)F(r)+r^2Q(r)=0,
\end{equation}
where, 
\begin{equation}
F(r)
=
\frac{r-4\pi r^3\left(\epsilon_{\rm tot}-P_{\rm tot}\right)}
{r-2M(r)},
\end{equation}
and
\begin{equation}
\begin{aligned}
Q(r)
={}&
\frac{4\pi r
\left[
5\epsilon_{\rm tot}
+9P_{\rm tot}
+\displaystyle\sum_i
\frac{\epsilon_i+P_i}{\partial P_i/\partial\epsilon_i}
-\frac{6}{4\pi r^2}
\right]}
{r-2M(r)}
-4\left[
\frac{M(r)+4\pi r^3P_{\rm tot}}
{r^2(-2M(r)/r)}
\right]^2.
\end{aligned}
\end{equation}

\section{Results and discussion} 
\label{sec:results}

In this section, we investigate the effects of the scalar-cut potential and DM on neutron star properties and observables. By solving Eqs.~(\ref{Eq:tov:P}) and (\ref{Eq:tov:M}), which govern the hydrostatic equilibrium and mass distribution of the two-fluid system, we obtain the mass–radius ($M$–$R$) relations and further solving Eq.(\ref{Eq:Lambda}), we obtain tidal deformability of DANS. We consider different DM fractions $f_{\rm DM}$, to examine how the presence of DM influences the properties of DANS. Also, we investigate the impact of the scalar-cut potential on the EOS and the resulting neutron star structure, both in the absence and presence of DM.

\begin{table}[h]
\centering
\renewcommand{\arraystretch}{1.8}
\setlength{\tabcolsep}{5pt}
\begin{tabular}{|c|c|c|c|c|c|c|c|}
\hline
$k_0$ & 4.570 & $k_1$ & 2.663 & $k_2$ & $-9.988$ & $k_3$ & $-4.320$ \\
\hline
$k_4$ & $-0.161$ & $g_4$ & 63.535 & $\rho_0$ (fm$^{-3}$) & 0.15 & $m_v$ (MeV) & 673.648 \\
\hline
$f_\pi$ (MeV) & 93 & $f_K$ (MeV) & 115 & $m_\pi$ (MeV) & 139 & $m_K$ (MeV) & 494 \\
\hline
$m_\omega$ (MeV) & 783 & $\sigma_0$ (MeV) & $-93$ & $\zeta_0$ (MeV) & $-96.874$ & $\chi_0$ (MeV) & 282 \\
\hline
$m_{\varphi_D}$ (MeV)  &  $1\times10^{-22}$ &
$m_{V_D}$ (MeV) & $10$ &
$M_p$ (MeV) & $1.22\times10^{22}$  &
-  & -\\
\hline
\end{tabular}
\caption{Various parameters utilized in the present calculation.}
\label{params}
\end{table}

 \subsection{Effects of scalar-cut potential on neutron star properties ($f_{\rm DM}=0$)}

 \begin{figure*}[t]
    \centering
    \includegraphics[width=0.99\textwidth]{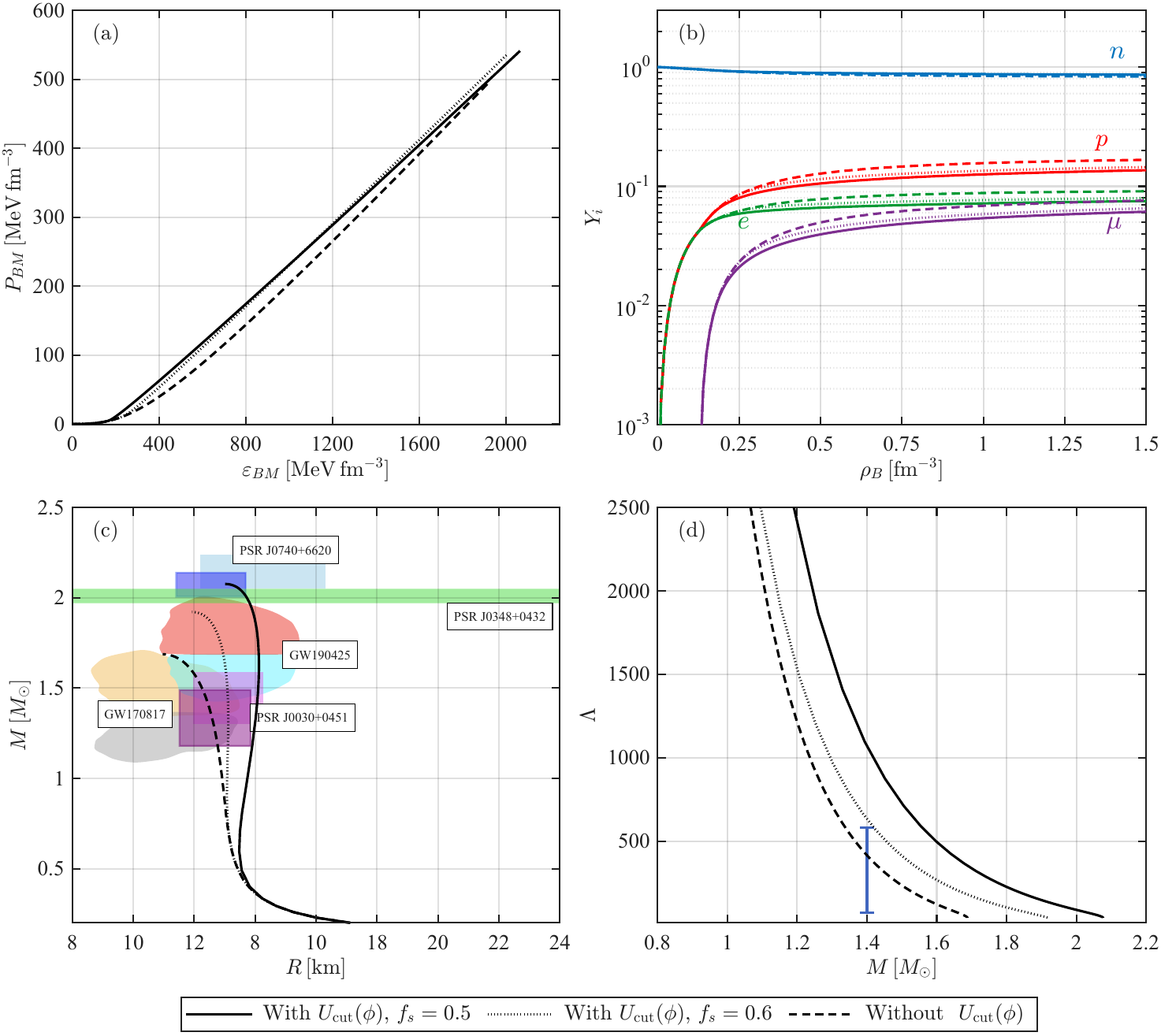}
    \caption{Baryonic equation of state and neutron star properties for $f_{\rm DM}=0$. The panel (a) shows the pressure as a function of energy density,  the panel (b) shows the particle fractions as functions of baryon density and panels (c) and (d) presents the $M-R$ and $\Lambda-M$ relations, respectively. Results are shown for two values of the parameter $f_s$ with the $U_{\rm cut}(\phi)$ ($f_s=0.5$ and $0.6$), and for the case without the $U_{\rm cut}(\phi)$.}
    \label{Fig:baryonic}
\end{figure*}

 Figure~\ref{Fig:baryonic} shows the effects of the $U_{\text{cut}}(\phi
)$ on the EOS, particle fraction ($Y_i$), $M-R$ and $\Lambda-M$ relations of neutron stars. The $U_{\text{cut}}(\phi
)$ given in Eq. (\ref{Eq:cutpotential}) modifies the EOS at high baryon densities, thereby influencing the macroscopic properties of 
neutron stars.  We compare the results obtained with the $U_{\text{cut}}(\phi
)$ for $f_s=0.5$ and $0.6$ with those obtained in its absence.  

In Fig.~\ref{Fig:baryonic} (a), we show the variation of the total baryonic pressure $P_{BM}$, as a function of the total baryonic energy density $\epsilon_{BM}$, calculated using Eq. (\ref{Eq:bm_pressure}). As observed from Fig.~\ref{Fig:baryonic} (a), when the density is low, the three curves are close to each
other and at $\epsilon_{BM}=162.495$ and $232.171$  $\mathrm{MeV\ fm}^{-3}$, the pressure starts to increase for $f_s=0.5$ and 0.6, respectively, from their value without $U_{\text{cut}}(\phi
)$. Therefore, the inclusion of the $U_{\text{cut}}(\phi
)$ increases the pressure at a given energy density, reflecting the influence of the modified scalar fields and effective Lagrangian on the EOS and thereby stiffening of the EOS. As $f_s$ increases from $0.5$ to $0.6$, the EOS initially softens, with $P_{BM}$ decreasing from $89.75$ to $81.8$ $\mathrm{MeV\ fm}^{-3}$ at $\epsilon_{BM}=500$ $\mathrm{MeV\ fm}^{-3}$ and as $\epsilon_{BM}$ reaches above $1049$ $\mathrm{MeV\ fm}^{-3}$, the EOS transitions to a stiffer behaviour, as evident from Fig.~\ref{Fig:baryonic} (a).
Figure~\ref{Fig:baryonic} (b) shows the particle fractions $Y_i$ of neutrons, protons, electrons, and muons as functions of the baryon density $(\rho_B)$ , which is given by
\begin{equation}
    Y_i=\frac{\rho_v^i}{\rho_B}.
\end{equation}
Here, $i=n,p,e$ and $\mu$. At low baryon densities, the matter is predominantly composed of neutrons, with smaller contributions from protons, electrons, and muons. As $\rho_B$ increases, the proton and lepton fractions increase, whereas the neutron fraction slowly decreases from 1, indicating changes in the composition of matter under the conditions of charge neutrality and $\beta$-equilibrium for all three configurations. As evident from Fig.~\ref{Fig:baryonic} (b), the inclusion of the $U_{\text{cut}}(\phi)$ decreases the particle fractions negligibly compared with the case without the potential. Furthermore, within $U_{\text{cut}}(\phi)$ potential, decreasing $f_s$ leads to a reduction in the particle fractions. The Fig.~\ref{Fig:baryonic} (c), shows the $M$–$R$ relation for neutron stars obtained with $U_{\text{cut}}(\phi)$ for $f_s=0.5$ and $0.6$, and without the $U_{\text{cut}}(\phi)$. The mass and radius are obtained by solving the TOV equations given in Eq. (\ref{Eq:tov:P}) and (\ref{Eq:tov:M}), keeping the $P_{DM}$ and $\epsilon_{DM}$ zero. As discussed earlier for Fig.\ref{Fig:baryonic} (a), the inclusion of the $U_{\text{cut}}(\phi)$ potential stiffens the EOS, thereby modifying the resulting $M$–$R$ relation. As shown in Fig.\ref{Fig:baryonic} (c), the maximum mass obtained without $U_{\text{cut}}(\phi)$ is $ 1.687\ M_\odot$, corresponding to a radius of $10.964$ km. In contrast, including $U_{\text{cut}}(\phi)$ increases the maximum mass (corresponding radius) to $2.076\ M_\odot$($13.018$ km) and $1.922\ M_\odot$($11.930$ km) for $f_s=0.5$ and $0.6$, respectively. 
We compare our results with observational constraints from massive pulsars, including PSR J$0348+0432$~\cite{Antoniadis:2013}, PSR J$0030+0451$~\cite{Miller:2019,Riley:2019}, and PSR J$0740+6620$~\cite{Miller:2021,Riley:2021,Fonseca:2021}, as well as gravitational wave observations of GW170817~\cite{Abbott:2018} and GW190425~\cite{Abbott:2020}. The $M$–$R$ curve obtained with $U_{\text{cut}}(\phi)$ for $f_s=0.5$ is in good agreement with all the considered observational constraints, Whereas, the curve obtained for $f_s=0.6$ and without $U_{\text{cut}}(\phi)$ is consistent with the only constraints from GW170817, GW190425, and PSR J$0030+0451$. 
We also studied the tidal deformability using the same EOS. The variation of the dimensionless tidal deformability $\Lambda$, calculated using Eq. (\ref{Eq:Lambda}), as a function of mass $M$ is shown in Fig.\ref{Fig:baryonic} (d) for the three configurations considered. As evident from the Fig.\ref{Fig:baryonic} (d), $\Lambda$ decreases with increasing stellar mass for all three cases. This behavior is associated with the increasing compactness of neutron stars, which makes them less susceptible to deformation under an external tidal field. The inclusion of the $U_{\mathrm{cut}}(\phi)$ potential enhances the tidal deformability, whereas increasing $f_s$, leads to a reduction in $\Lambda$, i.e, $\Lambda_{1.4}$ without the inclusion of $U_{\mathrm{cut}}(\phi)$ is $414.65$, it increases by value $653.85$ and $217.5$ for $f_s=0.5$ and 0.6, respectively. The observational constraint on $\Lambda_{1.4}$ associated with GW170817~\cite{Abbott:2018} is given in Fig.~\ref{Fig:baryonic} (d), providing a reference for assessing the consistency of our theoretical predictions with gravitational wave observations. The $\Lambda_{1.4}$ value obtained for the configuration without the potential, fall within the constrained region.

% Existing studies on the effects of the $\sigma$-cut potential in RMF models
% ~\cite{Maslov:2015,Zhang:2018,Wu:2020,Patra:2022,Ma:2022b} is consistent with  our studies
% on the scalar-cut potential.

% As in those works, the
% potential leaves the equation of state unchanged at low density and stiffens only at high densities, which raises the maximum mass from
% $1.687$ to $2.076$\,$M_\odot$ for $f_s=0.5$~\cite{Patra:2022,Zhang:2018}.
% The particle fractions $Y_i$ of the
% nucleons and leptons are negligibly affected, while the heavier species respond to the
% quenched scalar field~\cite{Patra:2022,Ma:2022b}.
% The radius and the tidal deformability of massive stars
% increases because of the stiffer equation of state provieded by the potential~\cite{Patra:2022,Ma:2022b}.

\subsection{Dark matter effects on neutron star structure and observables}

In this section, we investigate the influence of DM accumulation on neutron star properties in the presence of the $U_{\mathrm{cut}}(\phi)$. Specifically, we examine the effects of varying the DM mass $M_\chi$, and the interaction strength parameters, $ c_{V_D}=g_{V_D}/m_{V_D}$ and $ c_{\varphi_D}=g_{\varphi_D}/m_{\varphi_D}$, on the $M-R$ and $\Lambda-M$ relations of DANS. The scalar interaction $c_{\varphi_D}$ is attractive, whereas the vector interaction $c_{V_D}$ is repulsive.  To systematically explore these dependencies, we present three sets of comparisons, focusing separately on the variation of $M_\chi$,  $c_{V_D}$, and $c_{\varphi_D}$. For each case, we examine the corresponding changes in the $M$–$R$ curve and tidal deformability, thereby assessing their impact on the properties of DANS. The shaded regions show the constraints from PSR J0348+0432, PSR J0740+6620, PSR J0030+0451, GW170817 and GW190425. Numerical values are tabulated in Table ~\ref{tab:macroscopic_fs06}. 

\begin{figure*}[t]
    \centering
    \includegraphics[width=0.99\textwidth]{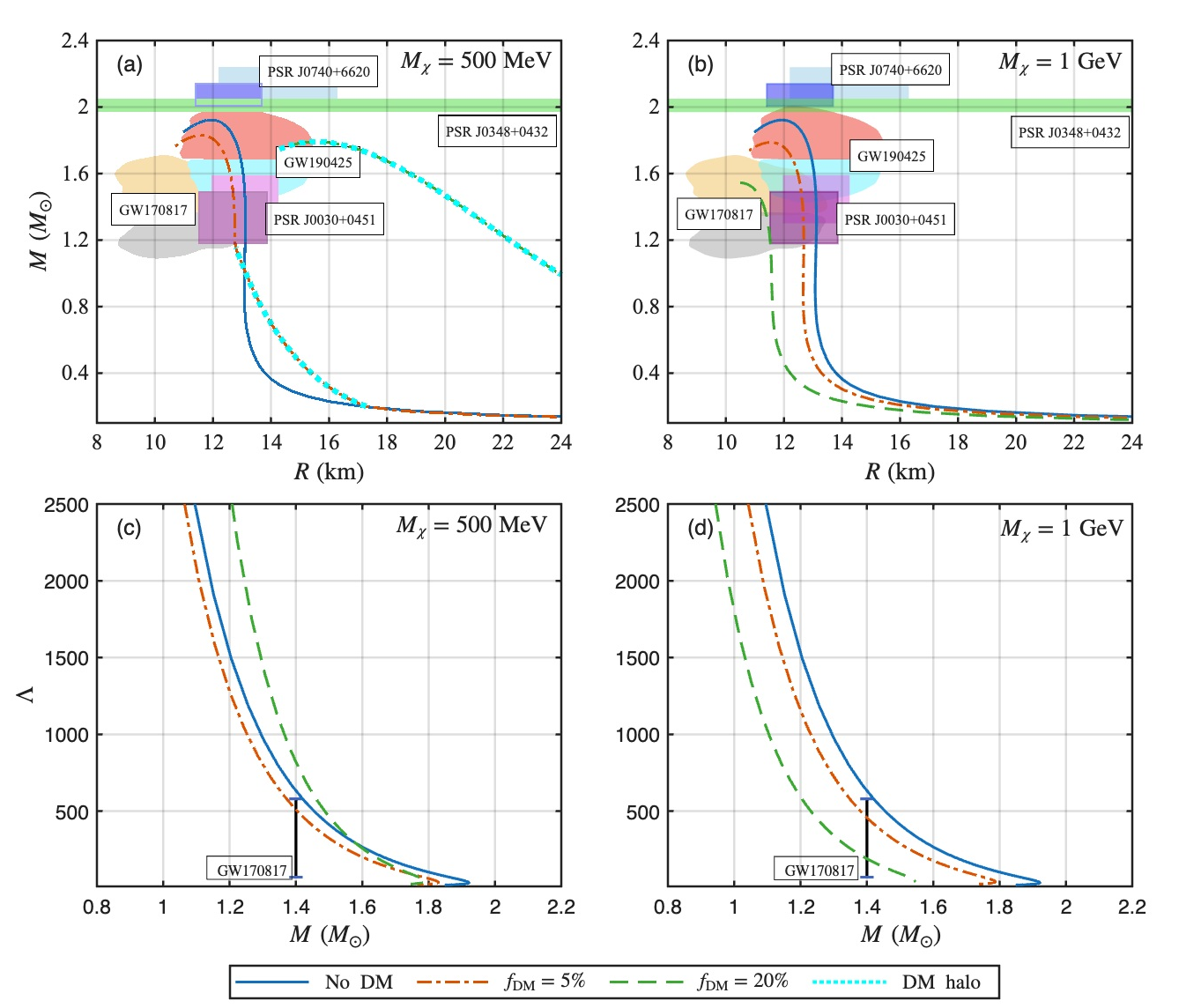}
    \caption{$M-R$ [panels (a) and (b)] and $\Lambda-M$ [panels (c) and (d)] relations for varying DM mass $M_\chi$ for $ c_{V_D}=10\ \rm GeV^{-1}$, $ c_{\varphi_D}=1\ \rm GeV^{-1}$ and $f_s=0.6$. Each curve corresponds to different $f_{\rm DM}$.}
    \label{fig:MR_mass}
\end{figure*}

\begin{figure*}[t]
    \centering
    \includegraphics[width=0.95\textwidth]{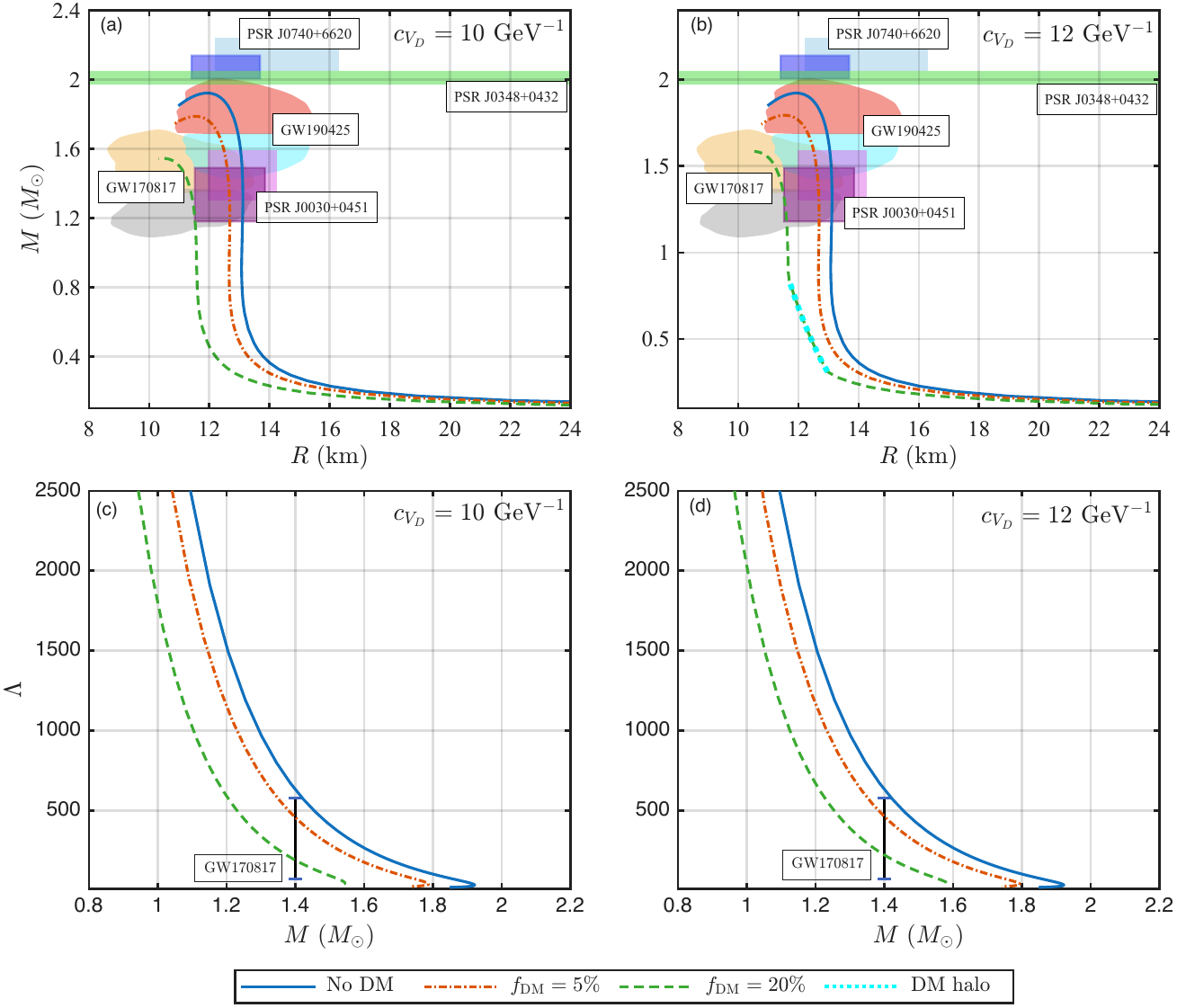}
    \caption{$M-R$ [panels (a) and (b)] and $\Lambda-M$ [panels (c) and (d)] relations for varying vector interaction parameter  $c_{V_D}$, for $M_\chi=1\ \rm GeV$, $ c_{\varphi_D}=1\ \rm GeV^{-1}$ and $f_s=0.6$. Each curve corresponds to different $f_{\rm DM}$.}
    \label{fig:MR_cv}
\end{figure*}

\begin{figure*}[t]
    \centering
    \includegraphics[width=0.99\textwidth]{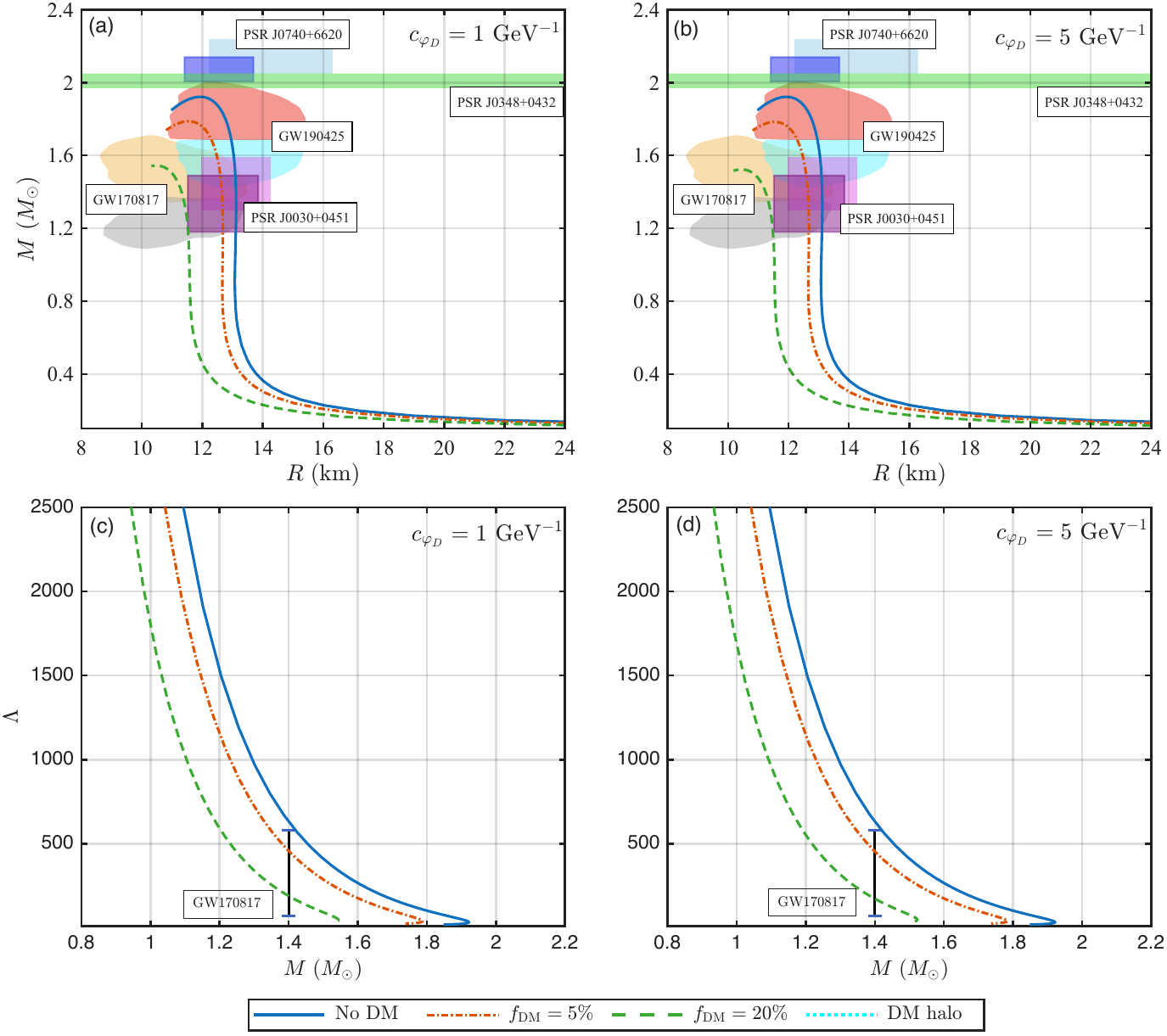}
    \caption{$M-R$ [panels (a) and (b)] and $\Lambda-M$ [panels (c) and (d)] relations for varying scalar interaction parameter $c_{\varphi_D}$, for $M_\chi=1\ \rm GeV$, $ c_{V_D}=10\ \rm GeV^{-1}$ and $f_s=0.6$. Each curve corresponds to different $f_{\rm DM}$.}
    \label{fig:MR_cphi}
\end{figure*}

Figure~\ref{fig:MR_mass} shows the $M-R$ and $\Lambda-M$ relations of DANS for varying $M_\chi$ for different DM fractions ($f_{\rm DM}=0$, $5$ and $20\%$) and scalar and vector interaction strength parameters $ c_{V_D}=10\ \rm GeV^{-1}$, $ c_{\varphi_D}=1\ \rm GeV^{-1}$, respectively, corresponding to $f_s=0.6$ of $U_{\rm cut}(\phi)$. As discussed earlier in Fig.\ref{Fig:baryonic} (c), the stellar mass and radius are obtained by solving Eqs. (\ref{Eq:tov:P}) and (\ref{Eq:tov:M}), respectively. The total radius is defined as $R=\max(R_{\rm BM},R_{\rm DM})$, therefore, configurations with $R_{\rm DM}<R_{\rm BM}$ correspond to DM core, whereas those with $R_{\rm DM}>R_{\rm BM}$ represent DM halo configurations. 

Figures~\ref{fig:MR_mass} (a) and (b) shows the $M-R$ relations for different DM mass $M_\chi= 500\ \rm MeV$ and $1\ \rm GeV$, respectively. 
As shown in Fig.~\ref{fig:MR_mass} (a), for $M_\chi=500$ MeV, the no DM case yields a maximum mass of $1.922\ \rm M_\odot$ with a corresponding radius of $11.93\ \rm km$ as discussed in earlier sections. Upon introducing DM, the maximum mass decreases to $1.829\ (1.790)\ \rm M_\odot$ for $f_{\rm DM}=5\ (20)\%$, with corresponding radii of $11.59\ (15.53)\ \rm km$, respectively. This reduction can be understood from the two-fluid TOV equations (Eqs.(\ref{Eq:tov:P}) and (\ref{Eq:tov:M})), where the gravitational field is determined by the combined mass and pressure of the BM and DM components. At a fixed $f_{\rm DM}$, the additional gravitational contribution from DM increases the gravitational attraction on the BM, resulting in a more compact configuration and higher central density. Since DM interacts with BM only through gravity, it does not provide direct pressure support to the baryonic component. Consequently, the presence of DM can lead to instability at a lower total gravitational mass than in the pure BM case. This effect becomes stronger as $f_{\rm DM}$ increases~\cite{Leung:2011, Ellis:2018, Karkevandi:2022} as evident from Fig.~\ref{fig:MR_mass} (a).  For $f_{\rm DM}=5\%$, halo configurations appear only for $R$ ranging between $12.77$--$17.27\ \rm km$, so the maximum mass configuration, with $R=11.59\ \rm km$ close to the no DM value, remains a DM core. For $f_{\rm DM}=20\%$, the larger DM content makes the halo persist up to the maximum mass point, giving maximum mass at $R=15.53\ \rm km$. Thus, increasing $f_{\rm DM}$ shifts the core-to-halo transition towards higher stellar mass. 
% , in agreement with the halo configurations reported in Refs.~\cite{Nelson:2019, Tolos:2015, Karkevandi:2022}.
% A light DM particle ($M_\chi=500$ MeV) has a more extended distribution, since the characteristic size of a free Fermi-gas DM star scales as $\sim M_{\rm Pl}/M_\chi^2$~\cite{Kouvaris:2015, Tolos:2015}, so halos form more readily than for heavier DM. The repulsive vector interaction pushes the DM outward, while the attractive scalar interaction pulls it inward.
As we increase the $M_\chi\ \rm to\ 1$ GeV as shown in Fig.~\ref{fig:MR_mass} (b), the maximum mass decreases to $1.787\ \rm and\ 1.544\ \rm M_\odot$ for $f_{\rm DM}=5\ \rm and\ 20\%$, with radii $11.549\ \rm to\ 10.471$ km. The decrease in maximum mass due to the presence of DM is stronger for $M_\chi=1$ GeV than for $M_\chi=500$ MeV, because heavier DM is more compact and has a lower Fermi pressure at a given number density. Heavier DM therefore settles in a core rather than a halo, and the radii remain close to the baryonic ones. This behaviour agrees with earlier studies, where lighter DM yields extended halos and heavier DM yields compact cores with smaller radii~\cite{Karkevandi:2022, Collier:2022, Nelson:2019}.
Figures~\ref{fig:MR_mass} (c) and (d) shows the $\Lambda-M$ relation for different $M_\chi$. In both cases, for $M_\chi=500\ \rm MeV\ and\ 1\ \rm GeV$, the $\Lambda$ decreases with increasing stellar mass as depicted in Fig.~\ref{fig:MR_mass} (c) and (d). From Eq. (\ref{Eq:Lambda}), since the $\Lambda$ is directly proportional to $(R/M)^5$ through the compactness, the halo configurations strongly enhance $\Lambda$ because of the larger radius, while DM cores reduce it. In Fig.~\ref{fig:MR_mass} (c), for $M_\chi=500$ MeV, we obtain $\Lambda_{1.4}=510.164$ and $820.744$ for $f_{\rm DM}=$ $5\%\ \rm and\ 20\%$ respectively, compared to the value $634.012$ for the no DM case.  With increase in $M_\chi$ to $1\ \rm GeV$ (Fig.~\ref{fig:MR_mass} (d)), the $\Lambda_{1.4}$ decreases to $458.173$ and $190.185$ for $f_{\rm DM}=5\%\ \rm and\ 20\%$ respectively. Comparing these values with the GW170817 constraint of $70\leq\Lambda_{1.4}\leq580$~\cite{Abbott:2018}, we find that the configurations with $M_\chi=500$ MeV and $f_{\rm DM}=20\%$, as well as the no DM configuration, lie above the upper bound, whereas the remaining configurations satisfy the constraint.

% The sensitivity of $\Lambda$ to DM mass and fraction is consistent with Refs.~\cite{Collier:2022, Karkevandi:2022, Nelson:201}, which point out that the $\Lambda$--$M$ relation can in principle discriminate DM halos from cores in future GW detections.

To study the effect of vector interaction parameter  $c_{V_D}$ on $M-R$ and $\Lambda-M$ relations, we fix $M_\chi\ \rm to\ 1\ \rm GeV$ and $ c_{\varphi_D}\ \rm to\ 1\ \rm GeV^{-1}$ and change  $c_{V_D}$ to $10$ and $12$ $\rm GeV^{-1}$ as depicted in Fig.~\ref{fig:MR_cv}. And they are compared with no DM case. The vector interaction is repulsive, so increasing  $c_{V_D}$ increases the $\epsilon_{DM}$ and $P_{DM}$ and stiffens the DM EOS~\cite{Nelson:2019, Mukhopadhyay:2016, Maselli:2017}. As observed in Fig.~\ref{fig:MR_cv} (a), for $ c_{V_D}=10\ \rm GeV^{-1}$ the maximum mass is $1.787\ (1.544)\ \rm M_\odot$ for $f_{\rm DM}=5\ (20)\%$, with radii $11.549\ (10.471)$ km. Increasing the coupling to $ c_{V_D}=12\ \rm GeV^{-1}$, Fig.~\ref{fig:MR_cv} (b), stiffens the DM EOS and thereby by increases the maximum mass to $1.794\ (1.584)\ \rm M_\odot$ and the radii to $11.552\ (10.533)$ km. The stiffer DM EOS also increases the extent of the DM distribution, therefore, in Fig.~\ref{fig:MR_cv} (b), for $f_{\rm DM}=20\%$, the halo is formed between radius $11.693-12.972\ \rm km$, but keeps the maximum mass as DM core.
% This is consistent with earlier studies of self-interacting DM, where repulsive interactions inflate the DM distribution and favour halo formation~\cite{Nelson:2019, Karkevandi:2022, Maselli:2017}.
The corresponding $\Lambda-M$ curves are shown in Figs.~\ref{fig:MR_cv} (c) and (d). As discussed earlier, larger radii of halo configurations enhance the tidal deformability, and the effect increases with  $c_{V_D}$. For $f_{\rm DM}=5\ (20)\%$ we obtain $\Lambda_{1.4}=458.173\ (190.185)$ for $ c_{V_D}=10\ \rm GeV^{-1}$ and as we increase  $c_{V_D}$ to $12\ \rm GeV^{-1}$, the $\Lambda_{1.4}$ increases to $463.606\ (222.192)$, consistent with the $\Lambda_{1.4}$ constraint~\cite{Abbott:2018}.

In Fig.~\ref{fig:MR_cphi}, we study the effect of scalar interaction term by fixing $ c_{V_D}=10\ \rm GeV^{-1}$ and $M_\chi=1\ \rm GeV$ and comparing the effect for $ c_{\varphi_D}=1$  (Figs.~\ref{fig:MR_cphi} (a) and (c)) and $5\ \rm GeV^{-1}$ (Figs.~\ref{fig:MR_cphi} (b) and (d)). The scalar interaction is attractive, so it lowers the DM energy per particle and softens the EOS~\cite{Nelson:2019, Mukhopadhyay:2016, Maselli:2017}. A softer DM EOS makes the DM more compact and pulls it towards the centre, so the trend is opposite to that of we discussed before for  $c_{V_D}$ in Fig.~\ref{fig:MR_cv}. For $ c_{\varphi_D}=1\ \rm GeV^{-1}$, in Fig.~\ref{fig:MR_cphi} (a), the maximum mass is $1.787\ (1.544)\ \rm M_\odot$ for $f_{\rm DM}=5\ (20)\%$ with radii $11.549\ (10.471)$ km.  
With increase in $c_{\varphi_D}$ from $1$ to $5\ \rm GeV^{-1}$ as shown in Fig.~\ref{fig:MR_cphi} (b), the mass and radius decreases to $1.784\ (1.523)\ \rm M_\odot$ and $11.538\ (10.446)$ km. Since the attraction binds the DM inside the baryonic surface, there is no halo formation in this case.
 % A similar softening and compaction of the DM component by attractive interactions has been reported in Refs.~\cite{Nelson:2019, Maselli:2017, Collier:2022}.
In the $\Lambda$--$M$ relation, depicted in Figs.~\ref{fig:MR_cphi} (c) and (d), the more compact DM reduces the radius and hence $\Lambda$. For $f_{\rm DM}=20\%$, $\Lambda_{1.4}$ changes from $190.185$ at $ c_{\varphi_D}=1\ \rm GeV^{-1}$ to $173.646$ at $ c_{\varphi_D}=5\ \rm GeV^{-1}$. Except for the no DM configuration, all other curves are consistent with the $\Lambda_{1.4}$ constraint from GW170817~\cite{Abbott:2018}.

\begin{figure*}[t]
    \centering    
    \includegraphics[width=1.0\textwidth]{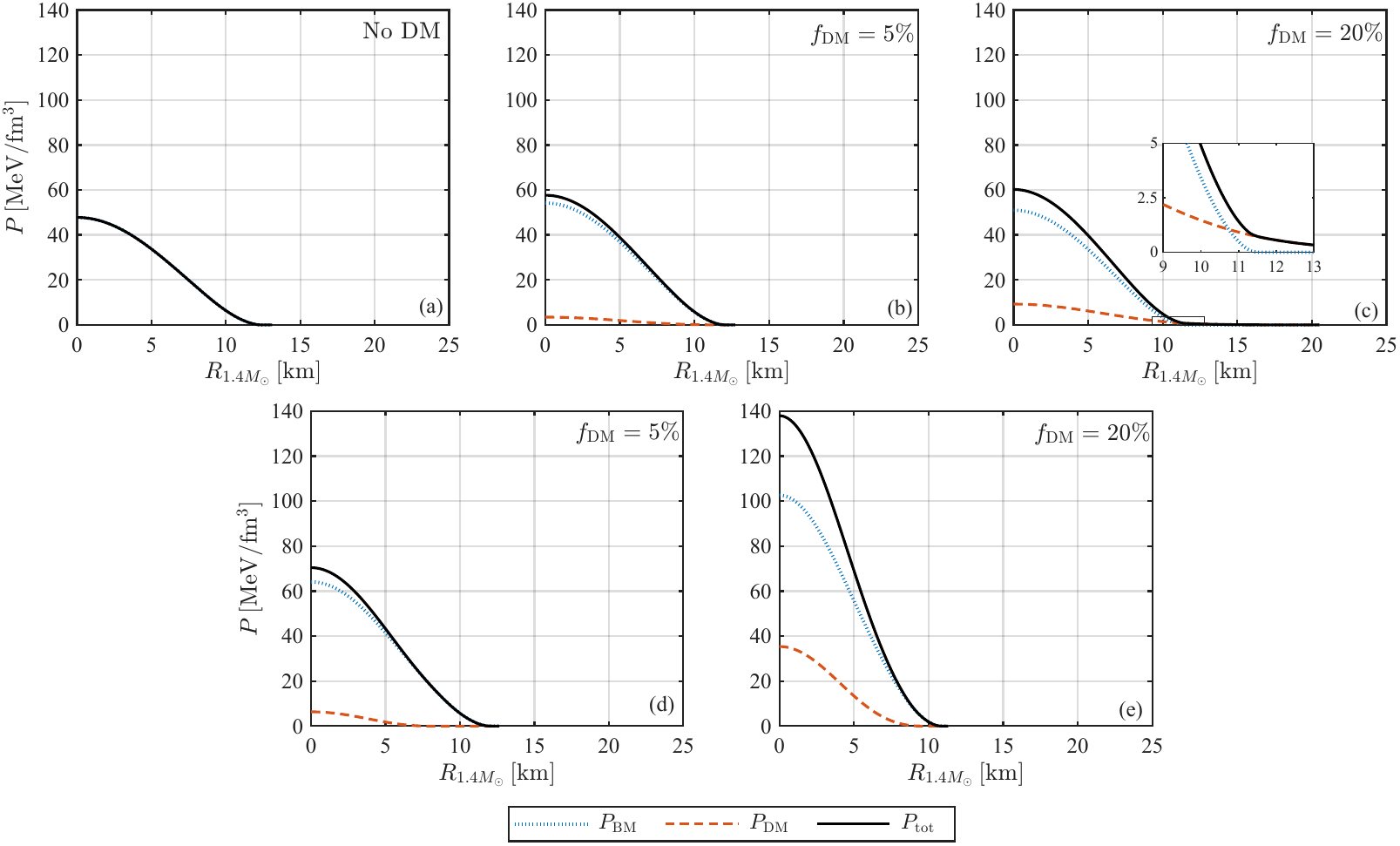}
    \caption{Radial pressure profiles of DANS for $1.4\ \rm M_\odot$ star as functions of radius for different $f_{\rm DM}=0$ [panel (a)], $5\%\ \rm and\ 20\%$ for $M_\chi=500\ \rm MeV$ [panels (b) and (c)] and $1\ \rm GeV$ [panels (d) and (e)]. The inset in panel (c) provides a magnified view of the pressure profiles near the stellar surface.}
    \label{Fig:pressure}
\end{figure*}

Figure~\ref{Fig:pressure} shows the radial pressure profiles of a $1.4\ \rm M_\odot$ star for $f_{\rm DM}=0$, $5$ and $20\%$ and for $M_\chi=500$ MeV and $1$ GeV, with $ c_{V_D}=10\ \rm GeV^{-1}$ and $ c_{\varphi_D}=1\ \rm GeV^{-1}$. As shown in Fig.~\ref{Fig:pressure}, $P_{\rm BM}$, $P_{\rm DM}$ and the total pressure $P_{\rm tot}=P_{\rm BM}+P_{\rm DM}$ are plotted separately. For $f_{\rm DM}=0\ (\rm no\ DM)$ in Fig.~\ref{Fig:pressure} (a), the star consists entirely of BM, so $P_{\rm tot}=P_{\rm BM}$. The central pressure is $47.78\ \rm MeV\,fm^{-3}$ and the pressure decreases monotonically to zero at the stellar surface, $R=13.11$ km. For $f_{\rm DM}=5\%$ and $M_\chi=500$ MeV as shown in Fig.~\ref{Fig:pressure} (b), the DM central pressure is small and has a value of $3.45\ \rm MeV\,fm^{-3}$, and it vanishes well inside the baryonic surface, so the DM forms a core ($R_{\rm DM}<R_{\rm BM}$) and the total radius, $R=12.73$ km, is set by the baryons. The DM contributes only $5.99\%$ of the central total pressure, but the central baryonic pressure rises to $54.19\ \rm MeV\,fm^{-3}$, compared with $47.78\ \rm MeV\,fm^{-3}$ for the pure baryonic star, giving $P_{\rm tot}=57.64\ \rm MeV\,fm^{-3}$. 
% As in the $1$ GeV case, this increase reflects the additional gravitational attraction of the DM, which compresses the baryonic matter~\cite{Ellis:2018, Karkevandi:2022}. The compression is, however, weaker than for $M_\chi=1$ GeV as ($P_{\rm BM}=64.07\ \rm MeV\,fm^{-3}$), because lighter DM is more diffuse, and a smaller fraction of its mass lies inside the region occupied by the baryons.
With increase in $f_{\rm DM}$ to $20\%$ as shown in Fig.~\ref{Fig:pressure} (c), the central DM pressure increases to $9.25\ \rm MeV\,fm^{-3}$, and the central baryonic and total pressures reach $50.96$ and $60.21\ \rm MeV\,fm^{-3}$, respectively. The halo formation region is shown in the inset in Fig.~\ref{Fig:pressure} (c). The baryonic pressure approaches zero at $R_{\rm BM}\approx11.5$ km, whereas the DM pressure continues to decrease slowly, out to $R_{\rm DM}=20.49$ km. For $R>R_{\rm BM}$ the total pressure is therefore supplied entirely by the DM, and a DM halo ($R_{\rm DM}>R_{\rm BM}$) extends beyond the baryonic surface.
As shown in Fig.~\ref{Fig:pressure} (d), increasing $M_\chi$ to $1$ GeV enhances the gravitational attraction associated with the DM component hence they remains confined within the neutron star for both DM fractions, leading to a stronger compression of the BM~\cite{Ellis:2018, Karkevandi:2022}. Consequently, for $f_{\rm DM}=5\%$, the central baryonic and DM pressures increase to $64.07$ and $6.33\ \rm MeV\ fm^{-3}$, respectively, compared with $54.19$ and $3.45\ \rm MeV\ fm^{-3}$ for $M_\chi=500$ MeV. Unlike the lighter DM case, the heavier DM is more centrally concentrated, resulting in a compact DM core rather than an extended halo as discussed earlier. 

\begin{figure*}[t]
    \centering
    \includegraphics[width=0.9\textwidth]{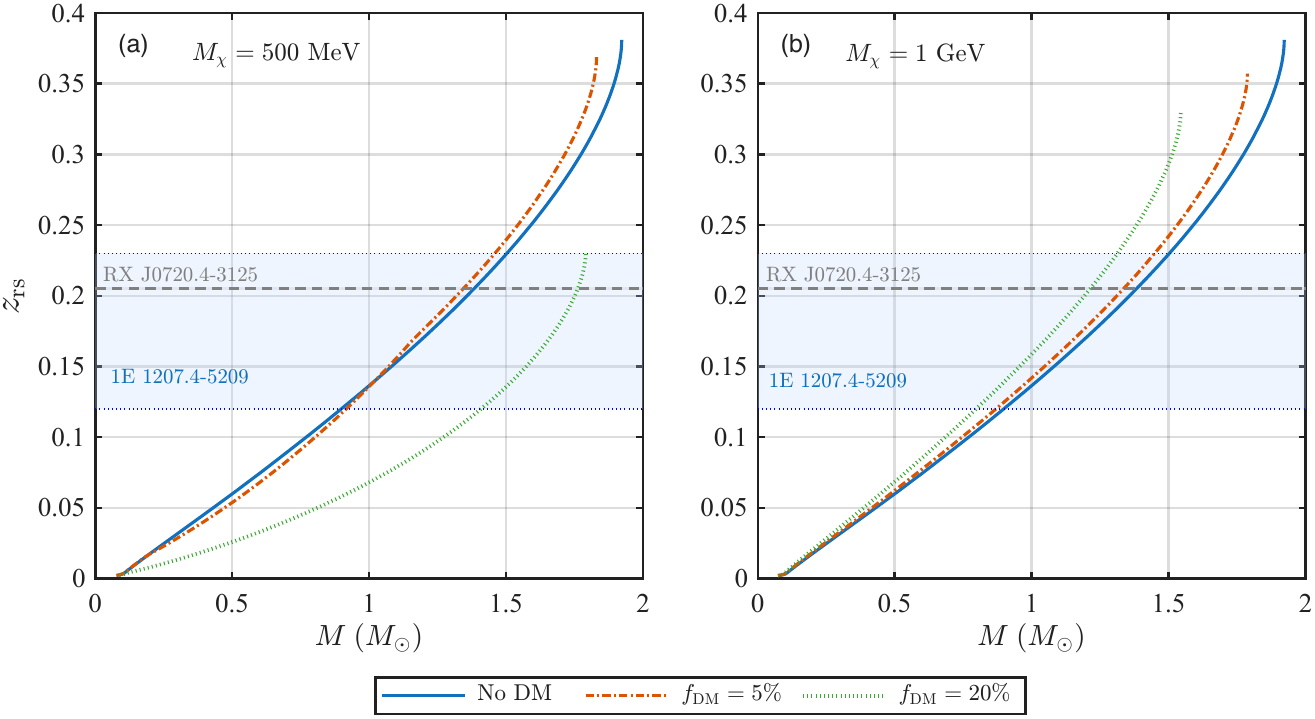}
    \caption{Gravitational redshift as a function of stellar mass for different $f_{\rm DM}$, considering $M_\chi=500$ MeV [panel (a)] and $M_\chi=1$ GeV [panel (b)] for $ c_{V_D}=10\ \mathrm{GeV}^{-1}$ and $ c_{\varphi_D}=1\ \mathrm{GeV}^{-1}$, with $f_s=0.6$.}
    \label{fig:redshift}
\end{figure*}

Figure~\ref{fig:redshift} shows the variation of gravitational redshift $z_{rs}$ with stellar mass for $M_\chi=500$ MeV and $1$ GeV, with $f_{\rm DM}=0$, $5\%$ and $20\%$. The gravitational redshift arises due to the loss of photon energy as it escapes the gravitational field of a compact star. 
Outside the star, the metric is determined by the total gravitational mass $M$, and for a photon emitted at radius $R$ the redshift is
\begin{equation}
z_{\rm rs}=\left(1-\frac{2M}{R}\right)^{-1/2}-1.
\label{eq:zrs}
\end{equation}
 % Such measurements exist for a few isolated neutron stars, notably 1E 1207.4-5209, for which absorption lines indicate $z_{\rm rs}\approx0.12$--$0.23$~\cite{Sanwal:2002}, and RX J0720.4-3125, for which $z_{\rm rs}=0.205$~\cite{Hambaryan:2017}.
In DANS, the DM changes both the total mass and the radius at a given central density, and therefore changes the $M-z_{\rm rs}$ relation. The shaded band marks the range inferred for 1E 1207.4-5209~\cite{Sanwal:2002} and RX J0720.4-3125~\cite{Hambaryan:2017}.

As observed in Fig.~\ref{fig:redshift} (a), for $M_\chi=500$ MeV, the no DM and $f_{\rm DM}=5\%$ curves remain close to each other. The $f_{\rm DM}=5\%$ curve lies slightly below the no DM curve for $M<1\ \rm M_\odot$, but shifts above it with the further increase in mass. The lower values at low mass are consistent with the DM halo configurations found for $f_{\rm DM}=5\%$, since a larger radius reduces the compactness, whereas the higher values as the mass increases reflect the compression by the DM core. The maximum redshift is $z_{\rm rs}=0.381$ for the no DM case ($1.922\ \rm M_\odot$) and $0.368$ for $f_{\rm DM}=5\%$ ($1.829\ \rm M_\odot$). In contrast, for $f_{\rm DM}=20\%$ the curve lies below the no DM and $f_{\rm DM}=5\%$ over the entire mass range, with $z_{\rm rs}=0.068$ at $M=1\ \rm M_\odot$ and $z_{\rm rs}=0.118$ at $M=1.4\ \rm M_\odot$, and it reaches only $z_{\rm rs}=0.231$ at the maximum mass ($M=1.790\ \rm M_\odot$). This strong reduction occurs because the halo increases the total radius and hence lowers the compactness at fixed mass. 
% The redshift of RX J0720.4-3125 is reached at $M\approx1.4\ \rm M_\odot$ for the no DM and $5\%$ cases, but only at $M\approx1.7\ \rm M_\odot$ for $f_{\rm DM}=20\%$.

As we increase the DM mass $M_\chi\ \rm to\ 1$ GeV as depicted in Fig.~\ref{fig:redshift} (b), the behaviour is reversed. The DM forms a core and compresses the star, so the radius at a given mass is smaller and $z_{\rm rs}$ is larger than the no DM case for both fractions, and the shift increases with $f_{\rm DM}$. At $M=1.4\ \rm M_\odot$, $z_{\rm rs}=0.255$ for $f_{\rm DM}=20\%$, compared with $0.208$ for the no DM case. The maximum redshifts are $0.357$ ($M=1.787\ \rm M_\odot$) and $0.330$ ($M=1.544\ \rm M_\odot$) for $f_{\rm DM}=5$ and $20\%$, respectively.

Recent studies of DM admixed stars in the two-fluid Cowling approximation, using the QMC-RMF4 model and four baryonic EOSs~\cite{Sotani:2025a,Sotani:2025b}, found that $M_{\max}$ decreases with increasing DM fraction and increases with repulsive coupling, as we find in Figs.~\ref{fig:MR_mass} and \ref{fig:MR_cv}. Including magnetic fields in the QMC-RMF4 framework revealed a possible core-to-halo transition for light DM~\cite{Routaray:2025}. Scalar and vector mediated DM interactions, with linear and quadratic scalar couplings, have been constrained using NICER and BNS observations~\cite{Grippa:2026}. They find that the repulsive vector coupling dominates over the attractive scalar one, with larger vector coupling giving larger $R$ and $\Lambda$ and a stronger scalar coupling giving more compact cores, which is closely related to our study of $c_{\varphi_D}$ and $c_{V_D}$. Numerical simulations of binary neutron star mergers with scalar DM show that the DM can form a persistent common cloud around the binary~\cite{Srikanth:2026}. This suggests that standard two-fluid tidal deformability estimates may not adequately describe such extended configurations, so $\Lambda$ for halos, such as our $M_\chi=500$ MeV case, should be interpreted with caution.

% requires \usepackage{multirow}
\begin{table*}[htbp]
\centering
\setlength{\tabcolsep}{5pt}
\renewcommand{\arraystretch}{1.5}

\begin{tabular}{|c|c|c|c|c|c|c|c|}
\hline
$M_\chi$
& $ c_{V_D}\ (\mathrm{GeV^{-1}})$
& $ c_{\varphi_D}\ (\mathrm{GeV^{-1}})$
& $f_{\rm DM}$ (\%)
& $M_{\max}\ (M_\odot)$
& $R_{\max}\ (\mathrm{km})$
& $\Lambda_{1.4}$
& $z_{\rm rs}$ \\
\hline
\multirow{3}{*}{500 MeV} & \multirow{3}{*}{10} & \multirow{3}{*}{1}
& 0  & 1.922 & 11.930 & 634.012 & 0.381 \\
\cline{4-8}
& & & 5  & 1.830 & 11.589 & 510.164 & 0.369 \\
\cline{4-8}
& & & 20 & 1.790 & 15.530 & 820.744 & 0.231 \\
\hline
\multirow{3}{*}{1 GeV} & \multirow{3}{*}{10} & \multirow{3}{*}{1}
& 0  & 1.922 & 11.930 & 634.012 & 0.381 \\
\cline{4-8}
& & & 5  & 1.787 & 11.549 & 458.173 & 0.357 \\
\cline{4-8}
& & & 20 & 1.544 & 10.471 & 190.185 & 0.331 \\
\hline
\multirow{3}{*}{1 GeV} & \multirow{3}{*}{12} & \multirow{3}{*}{1}
& 0  & 1.922 & 11.930 & 634.012 & 0.381 \\
\cline{4-8}
& & & 5  & 1.794 & 11.552 & 463.606 & 0.359 \\
\cline{4-8}
& & & 20 & 1.584 & 10.533 & 222.192 & 0.341 \\
\hline
\multirow{3}{*}{1 GeV} & \multirow{3}{*}{10} & \multirow{3}{*}{5}
& 0  & 1.922 & 11.930 & 634.012 & 0.381 \\
\cline{4-8}
& & & 5  & 1.784 & 11.538 & 454.376 & 0.357 \\
\cline{4-8}
& & & 20 & 1.523 & 10.446 & 173.646 & 0.325 \\
\hline
\end{tabular}
\caption{Macroscopic properties of DANS for different dark matter fractions, DM mass $M_\chi$ and interaction strengths $c_{\varphi_D}$ and $c_{V_D}$ , with $f_s=0.6$.}
\label{tab:macroscopic_fs06}
\end{table*}

\section{Summary}   
\label{sec:conclusion}

To summarize, we investigated the effects of the scalar-cut potential and fermionic dark matter (DM) on the macroscopic properties of neutron stars. We observed that $U_{\rm cut}(\phi)$ is essential for the baryonic EOS to support massive pulsars, as it stiffens the baryonic EOS at high density and raises the maximum mass from $1.687~M_\odot$ (without $U_{\rm cut}(\phi)$) to $2.076~M_\odot$ and $1.922~M_\odot$ for $f_s = 0.5$ and $0.6$, respectively. The inclusion of DM further modifies these properties such that it reduces the maximum mass, with the resulting configuration exhibiting either a DM core or a DM halo depending on the DM parameters. For lighter DM particles, the DM spreads around the baryonic matter, forming a halo and causing a moderate decrease in the maximum mass. In contrast, heavier DM particles are more concentrated at the centre, forming a core that leads to a greater reduction in the maximum mass and a smaller stellar radius. These effects become stronger as the DM fraction ($f_{\rm DM}$) increases.
The mediator interactions further influence the DM distribution and stellar properties. The repulsive vector interaction stiffens the DM EOS, so increasing  $c_{V_D}$ enhances the maximum mass and promotes a more extended DM distribution. In contrast, the scalar interaction is attractive and softens the DM EOS, so increasing $c_{\varphi_D}$ lowers the maximum mass and radius and suppresses halo formation. 
These structural differences are also reflected in the gravitational redshift, which decreases for halo configurations but increases for compact DM cores. The DM-induced modifications to stellar radius and tidal deformability can shift a given EOS towards or away from observational constraints. This degeneracy between the baryonic EOS and DM content is well known~\cite{Ellis:2018,Karkevandi:2022,Nelson:2019,Collier:2022}, suggesting that current observations may not exclude DM admixture but can constrain the allowed parameter space of $\{M_\chi,f_{\rm DM}, c_{V_D}, c_{\varphi_D}\}$.
Several extensions of the present work are possible. The mass, radius, tidal deformability, and gravitational redshift obtained here can be incorporated into Bayesian inference frameworks to constrain DM properties and nuclear matter parameters simultaneously. The DM induced modifications to tidal deformability and gravitational redshift can also be investigated through universal relations and radial and non-radial oscillation modes to explore possible deviations and the stability of two-fluid configurations. Further extensions include hyperons, quark phases, finite temperature effects, and protoneutron star evolution. Finally, future multimessenger observations and improved NICER radius measurements may provide tighter constraints on the dense matter EOS and the possible presence of DM inside compact stars.

 \section*{Acknowledgment}
A.K. sincerely acknowledge Anusandhan-National Research Foundation (ANRF), Government of India for funding of the research project under the Science and Engineering
Research Board-Core Research Grant (SERB-CRG) scheme (File No. CRG/2023/000557) for financial support.

\bibliographystyle{elsarticle-num} 
%\biboptions{numbers,sortcompress}
\bibliography{Ref}

\begin{thebibliography}{10}
\expandafter\ifx\csname url\endcsname\relax
  \def\url#1{\texttt{#1}}\fi
\expandafter\ifx\csname urlprefix\endcsname\relax\def\urlprefix{URL }\fi
\expandafter\ifx\csname href\endcsname\relax
  \def\href#1#2{#2} \def\path#1{#1}\fi

\bibitem{Rubin:1980}
V.~C. Rubin, W.~K.~J. Ford, N.~Thonnard, Astrophys. J. $\textbf{238}$ (1980)
  471.

\bibitem{Bosma:1981}
A.~Bosma, Astron. J. $\textbf{86}$ (1981) 1825.

\bibitem{Clowe:2006}
D.~Clowe, \textit{et al.}, Astrophys. J. Lett. $\textbf{648}$ (2006) L109.

\bibitem{Springel:2005}
V.~Springel, \textit{et al.}, Nature $\textbf{435}$ (2005) 629.

\bibitem{Ade:2016}
P.~A.~R. Ade, \textit{et al.}, Astron. Astrophys. $\textbf{594}$ (2016) A13.

\bibitem{Aghanim:2020}
N.~Aghanim, \textit{et al.}, Astron. Astrophys. $\textbf{641}$ (2020) A6.

\bibitem{Aprile:2018}
E.~Aprile, \textit{et al.}, Phys. Rev. Lett. $\textbf{121}$ (2018) 111302.

\bibitem{Aprile:2025}
E.~Aprile, \textit{et al.}, Phys. Rev. Lett. $\textbf{135}$ (2025) 221003.

\bibitem{Aalbers:2025}
J.~Aalbers, \textit{et al.}, Phys. Rev. Lett. $\textbf{135}$ (2025) 011802.

\bibitem{Bo:2025}
Z.~Bo, \textit{et al.}, Phys. Rev. Lett. $\textbf{134}$ (2025) 011805.

\bibitem{Kouvaris:2008}
C.~Kouvaris, Phys. Rev. D $\textbf{77}$ (2008) 023006.

\bibitem{Issifu:2025}
A.~Issifu, \textit{et al.}, Phys. Rev. D $\textbf{111}$ (2025) 083026.

\bibitem{Lopes:2025}
L.~L. Lopes, A.~Issifu, Phys. Dark Univ. $\textbf{48}$ (2025) 101922.

\bibitem{deLavallaz:2010}
A.~de~Lavallaz, M.~Fairbairn, Phys. Rev. D $\textbf{81}$ (2010) 123521.

\bibitem{Lattimer:2004}
J.~M. Lattimer, M.~Prakash, Science $\textbf{304}$ (2004) 536.

\bibitem{Ozel:2016}
F.~Ozel, P.~Freire, Annu. Rev. Astron. Astrophys. $\textbf{54}$ (2016) 401.

\bibitem{Baym:2018}
G.~Baym, \textit{et al.}, Rept. Prog. Phys. $\textbf{81}$ (2018) 056902.

\bibitem{Lattimer:2021}
J.~M. Lattimer, Annu. Rev. Nucl. Part. Sci. $\textbf{71}$ (2021) 433.

\bibitem{Busoni:2021}
G.~Busoni, Moscow Univ. Phys. Bull. $\textbf{77}$ (2022) 301.

\bibitem{Bell:2020}
N.~F. Bell, G.~Busoni, S.~Robles, M.~Virgato, JCAP $\textbf{2020}$ (2020) 028.

\bibitem{garani:19}
R.~Garani, J.~Heeck, Phys. Rev. D $\textbf{100}$ (2019) 035039.

\bibitem{Bertoni:2013}
B.~Bertoni, A.~Nelson, S.~Reddy, Phys. Rev. D $\textbf{88}$ (2013) 123505.

\bibitem{Bell:2023}
N.~F. Bell, G.~Busoni, S.~Robles, M.~Virgato, JCAP $\textbf{04}$ (2024) 006.

\bibitem{Bramante:2024}
J.~Bramante, N.~Raj, Phys. Rep. $\textbf{1052}$ (2024) 1.

\bibitem{Chandrasekhar:1939}
S.~Chandrasekhar, The University of Chicago Press, Chicago, Illinois, 1939.

\bibitem{Nattila:2022}
J.~Nättilä, J.~J.~E. Kajava, in: Handbook of X-ray and Gamma-ray
  Astrophysics, Springer, 2022, p.~11.

\bibitem{Das:2019}
A.~Das, T.~Malik, A.~C. Nayak, Phys. Rev. D \textbf{99} (2019) 043016.

\bibitem{Panotopoulos:2017}
G.~Panotopoulos, I.~Lopes, Phys. Rev. D \textbf{96} (2017) 083004.

\bibitem{Das:2022Galaxies}
H.~C. Das, A.~Kumar, B.~Kumar, S.~K. Patra, Galaxies \textbf{10} (2022) 14.

\bibitem{Lourenco:2022}
O.~Louren{\c{c}}o, C.~H. Lenzi, T.~Frederico, M.~Dutra, Phys. Rev. D
  \textbf{106} (2022) 043010.

\bibitem{Kumar:2024}
A.~Kumar, H.~Sotani, Phys. Rev. D \textbf{110} (2024) 063001.

\bibitem{Ellis:2018}
J.~Ellis, \textit{et al.}, Phys. Rev. D \textbf{97} (2018) 123007.

\bibitem{Sagun:2023}
V.~Sagun, \textit{et al.}, Astrophys. J. \textbf{958} (2023) 49.

\bibitem{Karkevandi:2022}
D.~Rafiei~Karkevandi, S.~Shakeri, V.~Sagun, O.~Ivanytskyi, Phys. Rev. D
  \textbf{105} (2022) 023001.

\bibitem{Karkevandi:2024}
D.~Rafiei~Karkevandi, M.~Shahrbaf, S.~Shakeri, S.~Typel, Particles \textbf{7}
  (2024) 201.

\bibitem{Sandin:2009}
F.~Sandin, P.~Ciarcelluti, Astropart. Phys. \textbf{32} (2009) 278.

\bibitem{Ciarcelluti:2011}
P.~Ciarcelluti, F.~Sandin, Phys. Lett. B \textbf{695} (2011) 19.

\bibitem{Hippert:2023}
M.~Hippert, \textit{et al.}, Phys. Rev. D \textbf{107} (2023) 115028.

\bibitem{Fornal:2018}
B.~Fornal, B.~Grinstein, Phys. Rev. Lett. \textbf{120} (2018) 191801.

\bibitem{McKeen:2018}
D.~McKeen, A.~E. Nelson, S.~Reddy, D.~Zhou, Phys. Rev. Lett. \textbf{121}
  (2018) 061802.

\bibitem{Motta:2018}
T.~F. Motta, P.~A.~M. Guichon, A.~W. Thomas, Int. J. Mod. Phys. A \textbf{33}
  (2018) 1844020.

\bibitem{Nelson:2019}
A.~E. Nelson, S.~Reddy, D.~Zhou, JCAP \textbf{07} (2019) 012.

\bibitem{Ivanytskyi:2020}
O.~Ivanytskyi, V.~Sagun, I.~Lopes, Phys. Rev. D \textbf{102} (2020) 063028.

\bibitem{Shawqi:2024}
S.~Shawqi, S.~M. Morsink, Astrophys. J. \textbf{975} (2024) 123.

\bibitem{Rutherford:2024}
N.~Rutherford, C.~Prescod-Weinstein, A.~Watts, Phys. Rev. D \textbf{111} (2025)
  123034.

\bibitem{Biesdorf:2024}
C.~Biesdorf, J.~Schaffner-Bielich, L.~Tolos, Phys. Rev. D \textbf{111} (2025)
  083038.

\bibitem{Leung:2011}
S.-C. Leung, M.-C. Chu, L.-M. Lin, Phys. Rev. D \textbf{84} (2011) 107301.

\bibitem{Hinderer:2008}
T.~Hinderer, Astrophys. J. \textbf{677} (2008) 1216.

\bibitem{Hinderer:2010}
T.~Hinderer, B.~D. Lackey, R.~N. Lang, J.~S. Read, Phys. Rev. D \textbf{81}
  (2010) 123016.

\bibitem{Das:2022}
A.~Das, T.~Malik, A.~C. Nayak, Phys. Rev. D $\textbf{105}$ (2022) 123034.

\bibitem{Leung:2022}
K.-L. Leung, M.-C. Chu, L.-M. Lin, Phys. Rev. D \textbf{105} (2022) 123010.

\bibitem{Xiang:2014}
Q.-F. Xiang, W.-Z. Jiang, D.-R. Zhang, R.-Y. Yang, Phys. Rev. C \textbf{89}
  (2014) 025803.

\bibitem{Grippa:2026}
F.~Grippa, G.~Lambiase, T.~K. Poddar, Phys. Rev. D \textbf{113} (2026) 063014.

\bibitem{Kumar:2025}
A.~Kumar, S.~Girmohanta, H.~Sotani, Eur. Phys. J. C \textbf{85} (2025).

\bibitem{Wang:2001}
P.~Wang, \textit{et al.}, Nucl. Phys. A \textbf{688} (2001) 791.

\bibitem{Wang:2005}
P.~t. Wang, Phys. Rev. C \textbf{72} (2005) 045801.

\bibitem{Maslov:2015a}
K.~A. Maslov, E.~E. Kolomeitsev, D.~N. Voskresensky, Phys. Lett. B \textbf{748}
  (2015) 369.

\bibitem{Maslov:2015}
K.~A. Maslov, E.~E. Kolomeitsev, D.~N. Voskresensky, Phys. Rev. C \textbf{92}
  (2015) 052801.

\bibitem{Maslov:2016}
K.~A. Maslov, E.~E. Kolomeitsev, D.~N. Voskresensky, Nucl. Phys. A \textbf{950}
  (2016) 64.

\bibitem{Kolomeitsev:2017}
E.~E. Kolomeitsev, K.~A. Maslov, D.~N. Voskresensky, Nucl. Phys. A \textbf{961}
  (2017) 106.

\bibitem{Wang:2001npa}
P.~Wang, \textit{et al.}, Nucl. Phys. A $\textbf{688}$ (2001) 791.

\bibitem{Papazoglou:1999prc}
P.~Papazoglou, \textit{et al.}, Phys. Rev. C $\textbf{59}$ (1999) 411.

\bibitem{Mishra2024Charmonium}
A.~Mishra, A.~Kumar, S.~P. Misra, Phys. Rev. D $\textbf{110}$ (2024) 014003.

\bibitem{Ping:2001ctp}
W.~Ping, Z.~Zong-Ye, Y.~You-Wen, Commun. Theor. Phys. $\textbf{36}$ (2001) 71.

\bibitem{Singh:2017cpc}
H.~Singh, A.~Kumar, H.~Dahiya, Chinese Phys. C $\textbf{41}$ (2017) 094104.

\bibitem{Pthakur:2024}
P.~Thakur, \textit{et al.}, Phys. Rev. C $\textbf{109}$ (2024) 025805.

\bibitem{Zhang:2018}
Y.~Zhang, J.~Hu, P.~Liu, Phys. Rev. C $\textbf{97}$ (2018) 015805.

\bibitem{Barik:2013prc}
N.~Barik, A.~Mishra, B.~K. Dash, Phys. Rev. C $\textbf{88}$ (2013) 015206.

\bibitem{Barik:1985prd}
N.~Barik, B.~K. Dash, Phys. Rev. D $\textbf{31}$ (1985) 7.

\bibitem{Kumari:2020mci}
M.~Kumari, A.~Kumar, Eur. Phys. J. Plus $\textbf{136}$ (2021) 19.

\bibitem{Kumar:2019cpc}
R.~Kumar, A.~Kumar, Chinese Phys. C $\textbf{43}$ (2019) 124109.

\bibitem{Kumar:2019epjc}
R.~Kumar, A.~Kumar, Eur. Phys. J. C $\textbf{79}$ (2019) 403.

\bibitem{narain:2006}
G.~Narain, J.~Schaffner-Bielich, I.~N. Mishustin, Phys. Rev. D $\textbf{74}$
  (2006) 063003.

\bibitem{Rutherford:2023}
N.~Rutherford, G.~Raaijmakers, C.~Prescod-Weinstein, A.~Watts, Phys. Rev. D
  $\textbf{107}$ (2023) 103051.

\bibitem{Flanagan:2008}
E.~E. Flanagan, T.~Hinderer, Phys. Rev. D $\textbf{77}$ (2008) 021502.

\bibitem{Damour:2009}
T.~Damour, A.~Nagar, Phys. Rev. D $\textbf{80}$ (2009) 084035.

\bibitem{Hinderer:2009}
T.~Hinderer, Astrophys. J. $\textbf{697}$ (2009) 964.

\bibitem{Antoniadis:2013}
J.~Antoniadis, \textit{et al.}, Science \textbf{340} (2013) 6131.

\bibitem{Miller:2019}
M.~C. Miller, \textit{et al.}, Astrophys. J. Lett. \textbf{887} (2019) L24.

\bibitem{Riley:2019}
T.~E. Riley, \textit{et al.}, Astrophys. J. Lett. \textbf{887} (2019) L21.

\bibitem{Miller:2021}
M.~C. Miller, \textit{et al.}, Astrophys. J. Lett. \textbf{918} (2021) L28.

\bibitem{Riley:2021}
T.~E. Riley, \textit{et al.}, Astrophys. J. Lett. \textbf{918} (2021) L27.

\bibitem{Fonseca:2021}
E.~Fonseca, \textit{et al.}, Astrophys. J. Lett. \textbf{915} (2021) L12.

\bibitem{Abbott:2018}
B.~P. Abbott, \textit{et al.}, Phys. Rev. Lett. \textbf{121} (2018) 161101.

\bibitem{Abbott:2020}
B.~P. Abbott, \textit{et al.} [LIGO~Scientific, Virgo], Astrophys. J. Lett.
  \textbf{892} (2020) L3.

\bibitem{Collier:2022}
M.~Collier, D.~Croon, R.~K. Leane, Phys. Rev. D \textbf{106} (2022) 123027.

\bibitem{Mukhopadhyay:2016}
P.~Mukhopadhyay, J.~Schaffner-Bielich, Phys. Rev. D \textbf{93} (2016) 083009.

\bibitem{Maselli:2017}
A.~Maselli, \textit{et al.}, Phys. Rev. D \textbf{96} (2017) 023005.

\bibitem{Sanwal:2002}
D.~Sanwal, G.~G. Pavlov, V.~E. Zavlin, M.~A. Teter, Astrophys. J. \textbf{574}
  (2002) L61.

\bibitem{Hambaryan:2017}
V.~Hambaryan, \textit{et al.}, Astron. Astrophys. \textbf{601} (2017) A108.

\bibitem{Sotani:2025a}
H.~Sotani, A.~Kumar, Phys. Rev. D \textbf{111} (2025) 123013.

\bibitem{Sotani:2025b}
H.~Sotani, A.~Kumar, Eur. Phys. J. C \textbf{85} (2025) 1438.

\bibitem{Routaray:2025}
P.~t. Routaray, Phys. Rev. D \textbf{111} (2025) 103045.

\bibitem{Srikanth:2026}
R.~Srikanth, T.~Dietrich, K.~Clough, Phys. Rev. D \textbf{113} (2026) 044045.

\end{thebibliography}

\section*{Appendix}
The mean-field equations governing the scalar fields ($\sigma$, $\zeta$ and $\delta$), the dilaton field ($\chi$) and the vector fields ($\omega$ and $\rho$), obtained from the thermodynamic potential using Eq. (\ref{Eq:mean_fields}), are given below. 

\begin{eqnarray}
k_0 \chi^2 \sigma 
- 4k_1 (\sigma^2 + \zeta^2 + \delta^2)\sigma 
- 2k_2 (\sigma^3 + 3\sigma \delta^2) 
- 2k_3 \chi \sigma \zeta \nonumber - \\\frac{d}{3}\chi^4 \left(\frac{2\sigma}{\sigma^2 - \delta^2}\right)+ \left(\frac{\chi}{\chi_0}\right)^2 m_\pi^2 f_\pi+U'_{\text{cut}}(\sigma) &= &
- \sum_j\frac{\partial M_j}{\partial \sigma} \nonumber \rho_s^j,
\end{eqnarray}
\begin{eqnarray}
k_0 \chi^2 \zeta 
- 4k_1 (\sigma^2 + \zeta^2 + \delta^2)\zeta 
- 4k_2 \zeta^3 
- k_3 \chi (\sigma^2 - \delta^2) \nonumber-\\ \frac{d}{3}\frac{\chi^4}{\zeta}
+ \left(\frac{\chi}{\chi_0}\right)^2 \left[
\sqrt{2} m_K^2 f_K - \frac{1}{\sqrt{2}} m_\pi^2 f_\pi
\right] +U'_{\text{cut}}(\zeta) &=& -\sum_j\frac{\partial M_j}{\partial \zeta} \nonumber \rho_s^j,
\end{eqnarray}
\begin{eqnarray}
k_0 \chi^2 \delta 
- 4k_1 (\sigma^2 + \zeta^2 + \delta^2)\delta 
- 2k_2 (\delta^3 + 3\sigma^2 \delta) 
+ 2k_3 \chi \delta \zeta \nonumber \\ + \frac{2}{3} d \chi^4 \left(\frac{\delta}{\sigma^2 - \delta^2}\right)+U'_{\text{cut}}(\delta) &=& 
-\sum_j\frac{\partial M_j}{\partial \delta} \nonumber \rho_s^j,
\end{eqnarray}
\begin{eqnarray}
k_0 \chi (\sigma^2 + \zeta^2 + \delta^2)
- k_3 (\sigma^2 - \delta^2)\zeta 
+ \chi^3 \left[1 + \ln\left(\frac{\chi^4}{\chi_0^4}\right)\right] 
+ 4k_4 \chi^3 \nonumber\\-\frac{4}{3} d \chi^3 \ln\left(
\frac{(\sigma^2 - \delta^2)\zeta}{\sigma_0^2 \zeta_0}
\right)  + \frac{2\chi}{\chi_0^2} \left[
m_\pi^2 f_\pi \sigma 
+ \left(\sqrt{2} m_K^2 f_K - \frac{1}{\sqrt{2}} m_\pi^2 f_\pi \right)\zeta
\right] \nonumber  \\-\frac{\chi}{\chi_0^2}(m_\omega^2 \omega^2 + m_\rho^2 \rho^2) &=&0,
\end{eqnarray}
\begin{eqnarray}
\frac{\chi^2}{\chi_0^2} m_\omega^2 \omega 
+ 4g_4 \omega^3 
+ 12g_4 \omega \rho^2 &=& 
 \sum g_\omega^j \rho_v^j,
\end{eqnarray}
\begin{eqnarray}
\frac{\chi^2}{\chi_0^2} m_\rho^2 \rho 
+ 4g_4 \rho^3 
+ 12g_4 \omega^2 \rho &=&  
\sum g_\rho^j\tau_3 \rho_v^j,
\end{eqnarray}
where $U'_{\text{cut}}(\phi)=\dfrac{\partial U_{\text{cut}}}{\partial\phi}$ is the derivative of the scalar-cut potential in Eq.(\ref{Eq:cutpotential}) with respect to the scalar field $\phi$, given by
\begin{equation}
U'_{\text{cut}}(\phi)=
\frac{\partial U_{\mathrm{cut}}}{\partial \phi}
=
-\frac{\alpha\bar\beta}{m_i}
\frac{
\exp\left[\bar\beta\left(1-\frac{m_i^*}{m_i}-f_s\right)\right]
}{
1+\exp\left[\bar\beta\left(1-\frac{m_i^*}{m_i}-f_s\right)\right]
}
\frac{\partial m_i^*}{\partial \phi}.
\end{equation}
\end{document}